\documentclass[aps,twocolumn,nofootinbib,preprintnumbers,superscriptaddress]{revtex4}

\usepackage[dvips]{color}
\usepackage[normalem]{ulem}
\usepackage{amsmath}
\usepackage{enumerate}
\usepackage{amsfonts}
\usepackage{yfonts}

\usepackage{subfigure}
\usepackage{psfrag}

\usepackage{epsfig}
\usepackage[latin1]{inputenc}
\usepackage{float}
\usepackage{graphicx}
\usepackage{cancel}
\usepackage{mathrsfs}
\usepackage{amssymb}
\usepackage{amsfonts}
\usepackage{amsmath}
\usepackage{slashed}

\usepackage{graphicx}
\usepackage{bm}

\def\({\left(}
\def\){\right)}
\def\[{\left[}
\def\]{\right]}
\def\<{\langle}
\def\>{\rangle}

\newcommand\half{{\ensuremath{\frac{1}{2}}}}
\newcommand\p{\ensuremath{\partial}}

\newcommand\field[1]{{\ensuremath{\mathbb{{#1}}}}}

\newcommand\vev[1]{{\ensuremath{\left\langle{#1}\right\rangle}}}

\newcommand{\RR}{\field{R}}

\newcommand{\be}{\begin{equation}}
\newcommand{\ee}{\end{equation}}
\newcommand{\bea}{\begin{eqnarray}}
\newcommand{\eea}{\end{eqnarray}}
\newcommand{\bwt}{\begin{widetext}}
\newcommand{\ewt}{\end{widetext}}

\newcommand{\bi}{\begin{itemize}}
\newcommand{\ei}{\end{itemize}}
\newcommand{\ben}{\begin{enumerate}}
\newcommand{\een}{\end{enumerate}}
\newcommand{\bca}{\begin{cases}}
\newcommand{\eca}{\end{cases}}
\newcommand{\bln}{\begin{align}}
\newcommand{\eln}{\end{align}}
\newcommand{\bst}{\begin{split}}
\newcommand{\est}{\end{split}}

\newcommand\al{{\alpha}}
\newcommand\ep{\epsilon}
\newcommand\sig{\sigma}

\newcommand\lam{\lambda}
\newcommand\Lam{\Lambda}
\newcommand\om{\omega}
\newcommand\Om{\Omega}

\newcommand\ga{{\ensuremath{{\gamma}}}}

\newcommand\De{{\ensuremath{{\Delta}}}}

\newcommand\nab{{\nabla}}

\newcommand\ov{\over}
\newcommand\ha{{\half}}

\def\le{\left}
\def\ri{\right}

\newcommand\sG{{\ensuremath{{\mathcal G}}}}
\newcommand\sH{{\ensuremath{{\mathcal H}}}}
\newcommand\sL{{\ensuremath{{\mathcal L}}}}
\newcommand\sM{{\ensuremath{{\mathcal M}}}}

\newcommand\sO{{\ensuremath{{\mathcal O}}}}
\newcommand\sR{{\ensuremath{{\mathcal R}}}}

\newcommand\sJ{{\mathcal J}}

\renewcommand{\Im}{\textrm{Im}\,}
\renewcommand{\Re}{\textrm{Re}\,}

\def\tg{\tilde g}

\begin{document}

\title {
Quantum phase transitions in holographic models of magnetism and superconductors}

\preprint{MIT-CTP 4126}

\author{Nabil Iqbal}
\affiliation{Center for Theoretical Physics, Massachusetts Institute of Technology,
Cambridge, MA 02139 }
\author{ Hong Liu}
\affiliation{Center for Theoretical Physics,
Massachusetts
Institute of Technology,
Cambridge, MA 02139 }
\author{M\'ark Mezei}
\affiliation{Center for Theoretical Physics, Massachusetts Institute of Technology,
Cambridge, MA 02139 }
\author{Qimiao Si}
\affiliation{Department of Physics and Astronomy,
Rice University,
Houston, TX 77005 }

\begin{abstract}

We study a holographic model realizing an ``antiferromagnetic'' phase in which a global $SU(2)$ symmetry representing spin is broken down to a $U(1)$ by the presence of a finite electric charge density. This involves the condensation of a neutral scalar field in a charged AdS black hole. We observe that the phase transition for both neutral and charged (as in the standard holographic superconductor) order parameters can be driven to zero temperature by a tuning of the UV conformal dimension of the order parameter, resulting in a quantum phase transition of the Berezinskii-Kosterlitz-Thouless type.
We also characterize the antiferromagnetic phase and an externally forced ferromagnetic phase by showing that they contain
the expected spin waves with linear and quadratic dispersions
respectively.

\end{abstract}

\today

\maketitle

\tableofcontents


\section{Introduction}

Quantum phase transitions naturally occur in strongly correlated many-body
systems, which often contain competing interactions and the concomitant competing orders.
When the transition is continuous, or first order with weak discontinuities,
it gives rise to fluctuations that are both quantum and
collective \cite{Hertz.76,Sachdev_book,Natphys.08}.
Such quantum criticality serves as a mechanism for some of the
most interesting phenomena in condensed matter physics, especially
in itinerant electronic systems \cite{Gegenwart.08,Lohneysen.07}.
Among these are the breakdown of Fermi liquid theory and the emergence
of unconventional superconductivity.

Quantum criticality is traditionally formulated within the Landau paradigm
of phase transitions. The critical theory expresses the fluctuations of the
order parameter, a coarse-grained classical variable manifesting the breaking
of a global symmetry, in $d+z$ dimensions \cite{Hertz.76}; here
$d$ is the spatial dimension, and $z$ the dynamic exponent.
More recent developments \cite{Si.01,Senthil.04}, however, have pointed to
new types of quantum critical points. New modes, which are
inherently quantum and are beyond order-parameter fluctuations,
emerge as part of the quantum critical excitations. Quantum criticality
is hence considerably richer and more delicate than its thermal classical
counterpart. In turn, new methods are needed to search for, study, and
characterize strongly coupled quantum critical systems.

\subsection{AdS$_2$ and the emergent IR CFT}
A promising new route has come from an unexpected source,
the AdS/CFT duality~\cite{AdS/CFT}, which
equates a gravity theory in a weakly curved $(d+1)$-dimensional
anti-de Sitter (AdS$_{d+1}$) spacetime with a strongly-coupled $d$-dimensional field theory living on its boundary.
The gravity is classical when the corresponding field theory
takes a large $N$ limit. This maps questions about strongly-coupled
many-body phenomena
to solvable single- or few-body classical problems in a curved geometry,
often that of a black hole.

For considering a boundary theory at a finite density, a particularly simple
gravity setup is a charged black hole in AdS~\cite{Romans:1991nq} which describes~\cite{Chamblin:1999tk} a boundary conformal field theory at a finite chemical potential for a $U(1)$ charge.
A rich body of phenomena has been found in this relatively simple context including
novel transport (see \cite{quantcritbh} for a review), holographic superconductors~\cite{holographicsc,hartnolletal}\footnote{Technically these should  be called charged superfluids, as the symmetry that is broken is global and is not gauged within the holographic framework.} (see \cite{Hartnoll:2009sz,herzogr,Horowitz:2010gk} for reviews), and non-Fermi liquids \cite{Lee09,Liu09,Cubrovic09,Faulkner09} (see also~\cite{Albash:2009wz,Basu:2009qz,soojong,chenkaowen,Faulkner:2009am,fabio,Denef09,hofman,Faulkner:2010tq}). In particular, at low energies the system
has an infrared fixed point described by a $(0+1)$-dimensional CFT \cite{Faulkner09}.
The CFT has nontrivial scaling behavior only in the temporal direction and is represented on the gravity side by a near horizon geometry AdS$_2\times \RR^{d-1}$.
This realization also offers a unified way, from the gravitational perspective, to understand both the onset of the superconducting instability and the emergence of non-Fermi liquids.

One finds that each operator in the UV theory develops an anomalous IR scaling dimension $\delta$. For example, if a scalar operator of charge $q$ and UV dimension $\De$ is dual to a minimally coupled scalar field in the bulk, its IR dimension is given by (see e.g. equation (43) of~\cite{Faulkner09}; in this paper we have set $g_F$ to $1$)
 \be \label{nuk}
 \delta = \ha + \sqrt{{\De(\De-d) \ov d (d-1)} - { q^2 \ov
 2 d (d-1)} + {1 \ov 4}} \ .
 \ee
This determines the onset of an instability: if $\delta$ becomes complex, the system is unstable\footnote{In the bulk, a complex $\delta$ is a violation of the AdS$_2$ BF bound and corresponds to an infinitely oscillating bulk field near the horizon and hence an instability, as pointed out in \cite{Denef:2009tp}.}. Due to the term proportional to $-q^2$ in~\eqref{nuk}, even an operator which is irrelevant in the UV can be unstable in the IR if $q$ is sufficiently large. This gives rise to interesting
examples of the phenomenon of dangerously irrelevant operators and possibly
suggests a novel type of pairing instability driven
by an IR fixed
point.\footnote{Note that here a clear understanding of the
underlying physical mechanism is somewhat hindered by the fact that
at this moment the gravity model provides only a macroscopic description of the condensation.}

A similar formula exists for fermions. Here there are no instabilities, but the IR dimension $\delta$ controls the nature of small excitations around a Fermi surface~\cite{Faulkner09} and in particular whether there exist stable quasi-particles. It also naturally yields fermion self-energies that are singular only in the temporal direction, properties that are characteristic of the
electron self-energy in models for non-Fermi
liquids~\cite{Varma89,Holstein:1973zz,Hertz.76}
and the spin self-energy of quantum critical heavy fermions \cite{Si.01}.

In this paper we extend these development to model
itinerant quantum magnetic systems, condensed matter
systems in which the quantum critical phenomenon is of great current
interest \cite{Gegenwart.08,Lohneysen.07,SachD}.

\subsection{Gravity formulation of a magnetic system}


In the low energy limit of an electronic system, spin-orbit couplings
become suppressed and spin rotations are decoupled from
spacetime rotations. Thus spin rotations remain a symmetry even though
the rotational symmetries may be broken by a lattice or other effects.
The low energy theory is then characterized by an $SU(2)$ global symmetry
which describes the spin rotation and a $U(1)$ symmetry describing the charge.
This $SU(2)$ symmetry
is of course only approximate and will be broken at high energies.
However, if one is only interested in universality classes describing
only low energy behavior (which does not involve spin-orbit couplings),
this high energy breaking will be irrelevant and one might as well replace it by
a UV completion in which the dynamics have an $SU(2)$ symmetry that is exact at all energies.

Using the standard AdS/CFT dictionary, the conserved currents $j^a_\mu, a=1,2,3$
for $SU(2)$ spin symmetry and $J^\mu$ for $U(1)$ charge
 should be dual to bulk gauge fields making up a $SU(2)_{\rm spin} \times U(1)_{\rm charge}$
gauge group in the bulk. For simplicity, in our bulk description we will consider
a theory in which this $SU(2) \times U(1)$ symmetry is exact to all energies.
Modeling spin-orbit couplings and understanding how to take into account
the relation with spacetime symmetries are interesting questions
which will be left for future study~(see also recent discussion in~\cite{Faulkner:2010tq}).

We will be interested in studying the gravity dual of an ``antiferromagnetic'' phase in a continuum limit. In such a limit the background value of the spin density is zero, but there exists a staggered spin order parameter $\Phi^a, a=1,2,3$ which transforms as a triplet under spin rotations. Its background value spontaneously breaks $SU(2)$ to the $U(1)$ subgroup corresponding to rotations about a single axis. This leads us to introduce a real
scalar field $\phi^a$ transforming as a triplet under $SU(2)_{\rm spin}$ (and neutral
under $U(1)_{\rm charge}$) as the corresponding bulk field. A phase with vanishing $SU(2)$ gauge fields but with a normalizable $\phi^a \neq 0$ in one direction can then be interpreted as an
antiferromagnetic (AFM) phase, or
a spin density wave (SDW) phase in an itinerant-electron context. A ``ferromagnet'' would have nonzero $SU(2)$ gauge fields, corresponding to a nonzero background spin density in the field theory; we also study this by applying a source analogous to an external magnetic field.

Note that the spirit of this discussion is parallel to that of holographic
superconductors in that the gravity description
captures the macroscopic dynamics of the order parameter and the symmetry breaking
pattern, but does not explain its microscopic origin.

\subsection{Quantum phase transitions in holographic models of symmetry breaking}

To describe the AFM order and its transition, we proceed in a way analogous to the
superconducting case.
The magnetic case considered here involves a neutral order parameter. Indeed, with the
exception of superconductivity, most ordering phenomena in condensed matter systems
involve a neutral order parameter; other examples include charge density wave order
and Pomeranchuck instability.
We are therefore led to consider holographic phase transitions involving condensation
of a neutral scalar field in a finite density system. Here we again consider
a charged black hole which is dual to a boundary conformal theory at a finite
chemical potential $\mu$ for a $U(1)$ charge.

Setting $q=0$ in~\eqref{nuk}
one finds that $\delta$ becomes complex if
\be \label{dobd}
{\rm max} \le({d -2 \ov 2}, d-\De_c \ri) < \De < \De_c \equiv {d + \sqrt{d} \ov 2}
 \ee
 where ${d -2\ov 2}$ corresponds to the unitarity bound on a scalar operator in $d$ spacetime dimensions.
On the gravity side this regime is where a scalar field $\phi$ satisfies the
Breitenlohner-Freedman (BF) bound~\cite{bf} of AdS$_{d+1}$ but violates the BF bound
of the near horizon AdS$_2$ region, as was recognized first in~\cite{hartlon}.

In this paper we construct the phase diagram for the condensation of such a neutral scalar field
in the $T-\De$ (equivalently, the $T-m^2$,
see Eq.~(18)) plane.
For all $\De$ in the range~\eqref{dobd} there exists a critical
temperature $T_c$ below which the neutral scalar field condenses. The phase transition
is second order with mean field exponents. For the standard quantization\footnote{The story
for alternative quantization
  is more involved, see discussion in sec.~\ref{sec:AFM} for details.} in AdS$_{d+1}$,
we find that $T_c$ decreases with an increasing $\De$ and approaches zero at $\De_c$.

Thus if we allow ourselves to vary $\De$ through $\De_c$, we find a quantum phase transition, as was previously noticed in \cite{hartlon, Denef:2009tp} in the charged case. Within the context of a specific boundary field theory varying the UV conformal dimension $\Delta$ seems somewhat artificial: however the physics here is actually controlled by the effective mass of the scalar in the near horizon AdS$_2$ region, and we later discuss several concrete ways to achieve this.
Interestingly, the quantum phase transition at $\De_c$ is not described by
mean field exponents, but is instead of the Berezinskii-Kosterlitz-Thouless (BKT) type with
an exponentially generated scale. More explicitly, for $\De \sim \De_c$, at $T=0$,
there exists an IR scale
 \be \label{bkts}
 \Lam_{IR} \sim \mu \exp \le(- {C \ov \sqrt{\De_c - \De}} \ri),
 \quad C = \pi \sqrt{d (d-1) \ov 2 \De_c - d}
 \ee
below which new physics appears (or in other words, the condensate becomes significant). From the point of view of the near horizon geometry AdS$_2$, this behavior can be understood
from the recent discussion in~\cite{soojong,Kaplan:2009kr} which argues that for a scalar field
in AdS with a mass square below the BF bound, such an exponential scale should generally
be generated.\footnote{At the BF bound an IR and UV fixed point, which corresponds
to standard and alternative quantization respectively, collide and move to the complex plane, in analogue with
the BKT transition. Note that in our case the standard and alternative quantization described here refer to those in AdS$_2$.} An explicit holographic model realizing this phenomenon was also recently constructed in \cite{Jensen:2010ga} and a different quantum phase transition which occurs outside the BF boundary (i.e. the BF bound of AdS$_2$ is not violated) has been constructed in~\cite{FHR}.

 \begin{figure}[h]
\begin{center}
\includegraphics[scale=0.55]{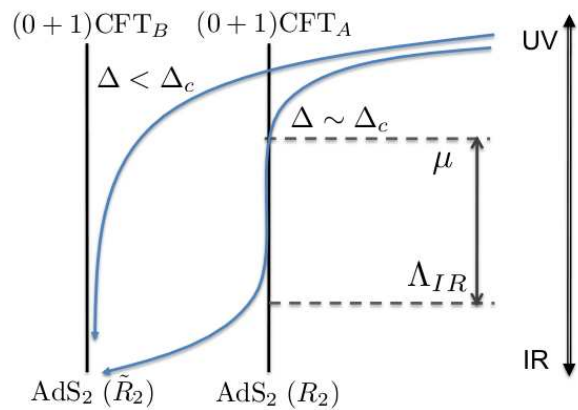}
\end{center}
\caption{A cartoon picture for the flow of the system induced by the condensation of a neutral scalar field. The CFT$_A$ refers to the $(0+1)$-dimensional IR CFT of the uncondensed system, described geometrically by an AdS$_2$ factor with radius $R_2$. When the dimension $\De$ of the operator is close to the quantum critical value $\De_c$, the system stays near this IR CFT for an exponentially long scale, before flowing to the new fixed point, $(0+1)$ CFT$_B$, described by an AdS$_2$ factor with a different radius $\tilde{R}_2$.}
\label{fig:rg}
\end{figure}

Below the IR scale $\Lambda_{IR}$ we show that the system in fact flows to a new IR fixed point which is controlled by a $(0+1)$-dimensional CFT,
dual to an AdS$_2$ with a different cosmological constant determined by the condensed vacuum of
the neutral scalar field. In Fig.~\ref{fig:rg} we give a cartoon picture of this flow.

The nature of the ordered phase can also be characterized in terms of its collective modes. The condensate breaks the global $SU(2) \to U(1)$: we show that the gapless spin waves expected from such a breaking
arise naturally in the gravity description, and that they obey the proper dispersion relations.

\begin{figure}[h]
\begin{center}
\includegraphics[scale=0.70]{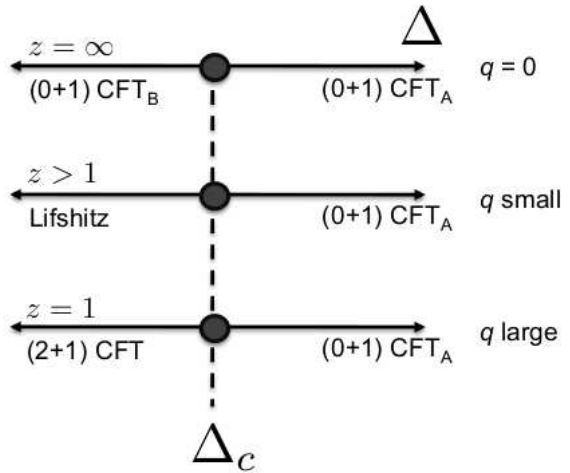}
\end{center}
\vskip -0.5cm
\caption{Tuning the UV dimension of the order parameter we find quantum phase transitions between a (0+1) dimensional IR CFT corresponding to AdS$_2$ in the unbroken phase and various types of symmetry-breaking phases. The type of symmetry-broken phase depends on the charge $q$ of the order parameter.}
\label{fig:qucr}
\end{figure}

The main focus of this current paper is on the condensation of a neutral scalar; however it is clear that one can immediately apply the above discussion to condensation of {\it charged} scalar operators
which give rise to the well-studied holographic superconductors. Equation~\eqref{nuk} implies that dialing
$\De$ one finds a quantum critical point at $\De_c$ given by the larger root of (for standard quantization)
 \be
 {\De_c (\De_c -d) \ov d (d-1)} - {q^2 \ov
 2 d (d-1)} + {1 \ov 4} = 0 \ .
 \ee
Again for $\De$ close to $\De_c$ the system will linger around the IR fixed point of
the uncondensed  phase for an exponentially large scale~\eqref{bkts} before settling
into new fixed points. In this case the gravity description of the
IR fixed points have been worked out before for $d=3$ in~\cite{Gubser:2009cg,Horowitz:2009ij,Gauntlett:2009dn,Gauntlett:2009bh} (see also \cite{Gubser:2008wz}). The stucture of the phase diagram will in general depend on the detailed form of the action, and non-minimal couplings such as those typically found in stringy embeddings (e.g. see \cite{Gauntlett:2009dn, Gauntlett:2009bh, Gubser:2009gp, Gubser:2009qm}) will change the results. It was found in \cite{Gubser:2009cg} that for minimally coupled scalar actions and potentials similar to ours and for $q$ small the system flows to a Lifshitz fixed point with a
dynamic exponent $z \sim {1 \ov q^2}$, while for larger values of $q$ the system flows to another AdS$_{4}$ with a different cosmological constant. Combined with the discussion above for the condensation of a neutral scalar field we thus find a unifying picture for the
quantum phase transitions for both charged and neutral order parameters as presented
in Fig.~\ref{fig:qucr}. Note that AdS$_2$ can be considered as describing a $d$-dimensional theory with a dynamic exponent $z = \infty$, while AdS$_{4}$ a theory
with $z = 1$.

Note that in our above discussion we have imagined the existence of an ``experimental knob'' which can be
used to adjust the UV scaling dimension $\De$ of an operator. It
is important to note that we do this purely for convenience, and the nature of our discussion is completely insensitive to the precise realization of such a knob. The most useful knob will likely depend on the UV geometry into which this AdS$_2$ is embedded. For example, if one is studying a holographic superconductor an external magnetic field will also allow one to tune the IR scaling dimension~\cite{Hartac}. A similar UV realization is provided in a D3/D5 brane construction in \cite{Jensen:2010ga}, where a precisely analogous transition is studied. For convenience in the remainder of this paper we will simply imagine that we are free to tune the bulk mass and thus directly the UV dimension of the scalar. In sec.~\ref{sec:IVA} we give a simple
model illustrating how this may be achieved.

A rough plan of the paper is as follows. In the next section we describe a holographic phase
transition corresponding to the condensation of a neutral scalar order parameter
in a finite density system in a probe approximation. In section~\ref{sec:back} we discuss
some qualitative features of the effects of backreaction and the zero temperature limit of our solution.
In section~\ref{sec:quan} we discuss the quantum phase transition that we find by tuning the UV conformal dimension. In section~\ref{sec:emb} we embed the scalar solution discussed in sec.~\ref{sec:AFM} into an $SU(2)$ system describing the spin rotational symmetry. We proceed to study the spin waves from spontaneous breaking
of the spin symmetry in the antiferromagnetic phase. In a probe limit we are able to isolate the spin wave excitations
directly in the bulk and find their dispersion relations.
We do not construct a spontaneous ferromagnet; however in section~\ref{sec:fmw} we do show that if one considers aligning the ``spins'' with an external magnetic field then techniques similar to those in section~\ref{sec:emb} can be used to find a spin wave which has a quadratic dispersion relation, in line with field theoretical expectations.
We conclude in section~\ref{sec:con} with a discussion of further directions.

\section{Condensation of a neutral order parameter at a finite density} \label{sec:AFM}

In this section we provide a gravity dual description of the condensation of a neutral
scalar field in a charged AdS black hole geometry which describes the onset of a real
 order parameter in the boundary theory at a finite density.
  We begin our analysis by studying the  transition at a finite temperature.
While our discussion applies to any spacetime dimension, for definiteness we will consider a $(2+1)$-dimensional boundary theory.

\subsection{Setup}

To put the system at a finite density we turn on a chemical potential $\mu$
for the $U(1)_{\rm charge}$ in the boundary. This is described on the gravity side by a charged black hole with a nonzero electric field for the corresponding
$U(1)$ gauge field $B_M$. The action for $B_M$ coupled to AdS gravity can be written as
 \be \label{grac}
 S = {1 \ov 2 \kappa^2} \int d^{4} x \,
 \sqrt{-g} \le[\sR +  { 6 \ov R^2} - R^2  G_{MN} G^{MN} \ri]
\ee
with $G_{MN} = \p_M B_N - \p_N B_M$ and $R$ is the
curvature radius of AdS. The equations of motion following from~\eqref{grac} are solved by the geometry of a charged black hole~\cite{Romans:1991nq,Chamblin:1999tk},
 \be \label{bhmetric1}
 {ds^2 \ov R^2} \equiv g_{MN} dx^M dx^N =  {r^2 } (-f dt^2 + d\vec x^2)  + {1\ov r^2} {dr^2 \ov f}
 \ee
 with
 \be \label{bhga2}
 f = 1 +  {3 \eta \ov  r^{4}} - {1+3\eta \ov r^3}, \qquad B_t = \mu \le(1- {1 \ov  r}\ri) \
 \ee
 where we have rescaled the coordinates so that the horizon is at $r=1$ and all coordinates are dimensionless (see Appendix~\ref{App:Aa} for more details). The chemical potential and temperature are given by
 \be \label{chem}
\mu \equiv  \sqrt{3} \eta^\ha, \qquad T = {3 \ov 4 \pi } \le(1 - \eta \ri) \ .
 \ee
$\eta$ is a parameter between $0$ and $1$, where $\eta =1$ corresponds to the extremal black hole with $T=0$ and $\eta =0$ corresponds to a finite temperature system with zero chemical potential (and charge density).\footnote{Note that since we are considering a conformal theory, only the dimensionless ratio ${\mu \ov T}$ is physically relevant, and it is clear that as $\eta$ is varied from $0$ to $1$, ${\mu \ov T}$ takes all values from $0$ to $\infty$.} In the zero temperature limit, the near horizon geometry
reduces to AdS$_2 \times \mathbb{R}^2$ with the curvature radius of the AdS$_2$ region
related to that of the UV $AdS_4$ by
\be \label{AdS2and4radii}
R_2 = {R \ov \sqrt{6}} \ .
\ee

As discussed in~\cite{hartlon,Faulkner09}, a neutral scalar field $\chi$ can develop an instability if the mass square of the scalar violates the near-horizon $AdS_2$ Breitenlohner-Freedman (BF) bound, while still satisfying the AdS$_{4}$ BF bound, i.e.~\footnote{For general $d$ boundary theory dimensions, equation~\eqref{tadkk} becomes
\be \label{tadkk}
 -{d^2 \ov 4} < m^2 R^2 < - {d (d-1) \ov 4} \ .
 \ee
 }
\be
-\frac{9}{4} < m^2 R^2  < -\frac{3}{2} \ . \label{instcrit}
\ee
where the lower limit is the BF bound in AdS$_{4}$, the upper limit
is the BF bound for the near horizon AdS$_2$ region, and we have used \eqref{AdS2and4radii} to convert from AdS$_2$ to AdS$_4$ radii.

Once this condition is met, the scalar will want to condense near the horizon but will be stable at infinity. The condensed solution will involve a nontrivial radial profile for the scalar; we will see that at low temperatures the scalar will probe the extreme values of its potential and nonlinearities in the potential will be important. We choose to study a nonlinear Mexican hat potential with the Lagrangian for $\chi$ given by
 \be \label{newp}
\sL_\chi = {1 \ov 2 \kappa^2 \lam} \le[- {1 \ov 2} (\p \chi)^2 - V(\chi) \ri]
 \ee
with
\be \label{expot}
 V(\chi) = {1 \ov 4R^2}  \le(\chi^2 + {m^2 R^2} \ri)^2
- {m^4 R^2 \ov 4 } \ .
\ee
Here $m^2$ is the effective mass near the point $\chi = 0$, and should be chosen to satisfy the condition \eqref{instcrit} (in particular, it is \emph{negative}). $\lam$ is a coupling constant and we have chosen the constant in~\eqref{newp} so that at $\chi=0$, there is no net contribution to the cosmological constant. The precise form of the potential in~\eqref{expot} is not important for our discussion below, provided that it does have a minimum and satisfies the condition \eqref{instcrit}\footnote{The condensed phase $q = 0$ solutions without the stabilizing $\chi^4$ term in the potential were studied in \cite{hartlon,Horowitz:2009ij}; at nonzero temperature these are similar to ours, but in the low temperature limit the structure described in Section \ref{sec:back} -- which depends critically on the existence of a minimum to the potential -- is not shared by those examples.}.

At zero temperature, we expect the scalar to condense until the value at the horizon reaches some point near the bottom of the Mexican hat, at which point the ``effective $AdS_2$ mass'' will again satisfy the $AdS_2$ BF bound and condensation will halt. At finite temperature, we expect a phase transition at some temperature $T_c$ below which $\chi$ condenses.
Note that at the classical level the coefficient of the $\chi^4$ term is arbitrary, as it can be absorbed into $\lambda$ in~\eqref{newp} via a rescaling of $\chi$.

\subsection{Phase diagram} \label{sec:phase}

We now seek the endpoint of the instability, i.e a nontrivial scalar profile for $\chi$.
We first consider the finite temperature case and take $\lam$ to be parametrically large
so that we can ignore the backreaction of $\chi$ to the background geometry. We will discuss the backreaction in section~\ref{sec:back} and there we argue that this approximation
is good even at zero temperature.

The equation of motion for $\chi(r)$ is given by
 \be
{1 \ov r^2} \p_r \le(r^4 f \p_r \chi \ri) -  \chi (\chi^2 + m^2R^2) = 0.
 \label{vve}
\ee
We are interested in a solution which is regular at the horizon and normalizable at the boundary, which can be found numerically.

We first consider the asymptotic behavior for $\chi$ near the horizon and the boundary. We require the solution to be regular at the horizon, i.e. to have the expansion
 \be
 \chi (r) = \chi_h + \chi_h' (r-1) + \cdots , \qquad r \approx 1
 \ee
At finite temperature the factor $f$ in~\eqref{vve} has a first order zero at the horizon, i.e.
$f(r) = 4 \pi T (r-1) + \cdots$. Demanding that \eqref{vve} be nonsingular then leads to a condition linking the near-horizon value of $\chi$ to its derivative:
 \be \label{1std}
 \chi'_h = {1 \ov 4 \pi T} \chi_h (\chi_h^2 + m^2 R^2 ) \ .
 \ee
A choice of $\chi_h$ fixes also $\chi_h'$ and thus completely specifies the solution.

 Near the boundary $r\to \infty$, the linearized equation of~\eqref{vve} gives the standard asymptotic behavior
 \be
\chi(r) \approx A r^{\Delta - 3} + B r^{-\Delta} , \qquad r \to \infty \label{scalasymp}
\ee
with $\De$ given by
\be
\Delta = \frac{3}{2} + \sqrt{m^2 R^2 + \frac{9}{4}} \ .
\ee
In the standard quantization $A$ has the interpretation of the source\footnote{For example, in section~\ref{sec:emb} we interpret $\chi$ as corresponding to the staggered magnetization, in which case $A$ can be interpreted as the staggered magnetic field.} while $B$ gives the response $\langle \Phi \rangle_A$ of the order parameter $\Phi$ dual to $\chi$
in the presence of source $A$.  Note that the mass range~\eqref{instcrit} lies within the range $-{9 \ov 4} < m^2R^2 < -{5 \ov 4}$ for which an alternative quantization exists, in which the roles of $A$ and $B$ are exchanged \cite{klebanov:1999}.
The conformal dimension of $\Phi$ (i.e. the ``dimension'' of fluctuations about the point $\chi = 0$) is given by $\De$ (standard quantization) and $3 - \De$ (alternative quantization) respectively.

\begin{figure}[!h]
\subfigure[$\;$$A$ and $B$ as a function of the horizon value of the scalar field $\chi_h$.]{\includegraphics[width=6cm]{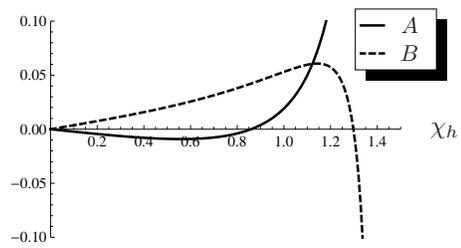}}\hspace{1.5cm}
\subfigure[$\;B$ as a function of $A$: different points on the curve correspond to different values of $\chi_h$.]{\includegraphics[width=6cm]{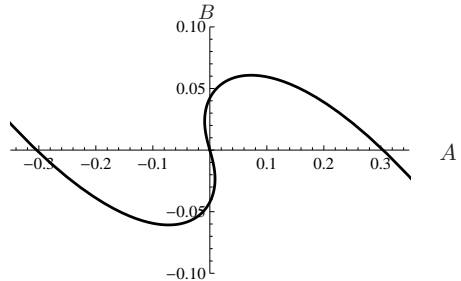}}
\caption{The constants $A$ and $B$ determine the behavior of the scalar profile asymptotically. This is a representative plot where we scan the case $m^2 R^2=-2.1$ and $T=0.00024$ (with $T/T_c=0.22$) by varying $\chi_h$. There is symmetry breaking if $A=0$ $B\neq0$ in the normal quantization or if $A\neq0$ $B=0$ in the alternative quantization.}
\label{sampleplot}
\end{figure}

The condensed phase for standard quantization is characterized by a normalizable nontrivial solution with $A =0$, and then $B$ gives the expectation value of the order parameter $\Phi$. For the alternative quantization we look instead for a solution with $B=0$, and $A$ then gives the expectation value. The task before us now is to pick a value for $\chi_h$, and then numerically integrate the radial evolution equation to the boundary. As expected for a sufficiently low temperature, we find
a nontrivial scalar hair solution, which means that there will exist a $\chi_h$ for which $A$ or $B$ vanishes, as shown in Figure~\ref{sampleplot}.

\begin{figure}[h]
\begin{center}
\includegraphics[scale=1.0]{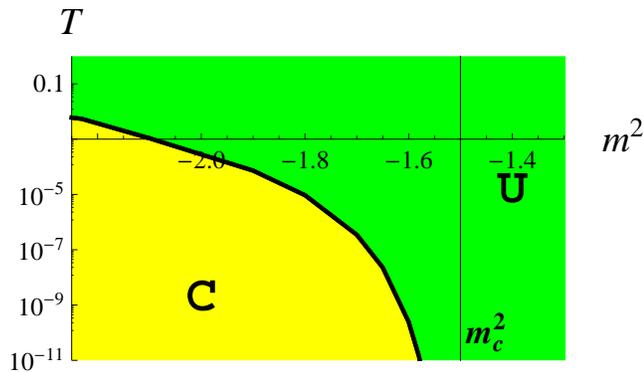}
\end{center}
\caption{Phase diagram for the standard quantization. Note logarithmic scale for $T$. $C$ denotes the condensed phase and $U$ the uncondensed phase. $T_c \to 0$ as $m^2 \to m_c^2$, leading to a quantum critical point.}
\label{fig:phaseA}
\end{figure}

For the standard quantization we find a continuous phase transition for all values of $m^2$ falling into the range~\eqref{instcrit}; there is a nontrivial profile for $\chi$ for $T$
smaller than some temperature $T_c$ and none for $T$ greater. In particular, as we increase $m^2$, the critical temperature $T_c$ decreases to zero as the upper bound in~\eqref{instcrit} is approached. Precisely at the critical value $m_c^2 R^2 = -{3 \ov 2}$, we find a quantum phase transition: the physics in the vicinity of this point is discussed in Sec.~\ref{sec:quan}. The phase diagram for standard quantization is plotted in
Fig.~\ref{fig:phaseA}.

For the alternative quantization, the phase structure in the vicinity of the quantum critical point at $m^2 = m_c^2$ is the same as in the standard quantization. However the global structure of the phase diagram is somewhat different; in particular the specific value of the mass $m^2R^2 = -\frac{27}{16}$ plays an important role. This value of the mass corresponds to the UV scaling dimension for which the first nonlinear (i.e. arising from the $\phi^4$ term in the potential) correction to the near-boundary asymptotics becomes degenerate with the term proportional to $B$. For $m^2 R^2 < -{27 \ov 16}$ the phase structure is as described above, but for $m^2 R^2 > -{27 \ov 16}$  one finds a new condensed phase in the {\it high temperature} regime. In particular, the critical temperature $T_{c2}$ appears to increase with $m^2$. These solutions appear to be closely related to the thermodynamically unstable scalar hair solutions constructed in \cite{Hertog:2004dr,Hertog:2004bb}, which studied uncharged black holes and thus correspond to the high-temperature limit of our construction. As these new phases appear to involve UV physics and are not related to the low temperature quantum-critical behavior that is the focus of this work, we defer an in-depth study of these phases and the critical value $m^2 R^2 = -{27 \ov 16}$ to later work.

\subsection{Critical exponents}

A continuous phase transition can be characterized by various critical exponents: in the normal quantization with $A$ the source and $B \sim \langle\Phi\rangle$ the expectation value we have the following behavior close to $T_c$:
\ben

\item The expectation value $\langle\Phi\rangle \propto (T_c - T)^\beta$ with mean field value $\beta = \ha$.

\item The specific heat:  $C \propto |T_c - T|^{-\al}$ with mean field value $\al=0$.

\item Zero field susceptibility: ${\p \langle \Phi\rangle \ov \p  A}|_{A=0} \propto |T-T_c|^{-\ga}$ with mean field value $\ga = 1$.

\item Precisely at $T=T_c$, $\langle\Phi\rangle \sim A^{1 \ov \delta}$ with mean field value $\delta =3$

\begin{figure}[h]
\begin{center}
\includegraphics[scale=0.75]{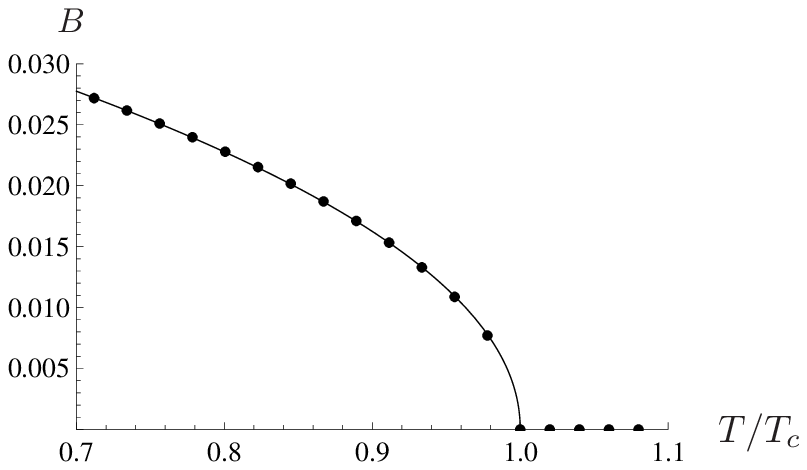}
\hskip0.5in
\includegraphics[scale=0.75]{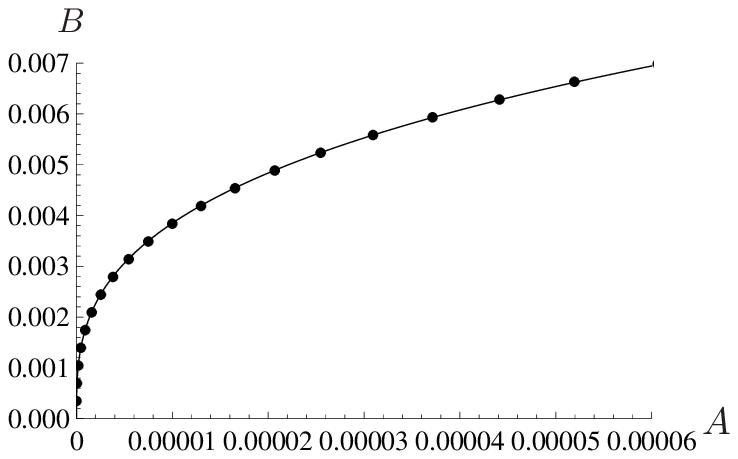}
\end{center}
\caption{
{\it Top:}
Plot for exponent $\beta$, defined as $B \sim (T_c-T)^{\beta}$. $\beta=0.49 \pm 0.03$ from numerical fit, compared with $\beta_{\mathrm{mean\;field}} = \frac{1}{2}$.
{\it Bottom:}
 Plot for exponent $\delta$, defined as $B \sim A^{\frac{1}{\delta}}$ at $T_c$. $\delta=3.03 \pm 0.05$ from numerical fit, compared with $\delta_{\mathrm{mean\;field}} = 3$. }
\label{fig:expB}
\end{figure}

\een
There are also other exponents associated with correlation functions at finite spatial or time separation which we will leave for future study. For our phase transition we find that all of the above exponents are precisely those of mean field theory. See fig.~\ref{fig:expB} for numerical fits of exponent $\beta$ and $\delta$.

The existence of mean field exponents in this classical gravity analysis is not surprising\footnote{Mean field exponents also appear in the phase transition associated with a charged scalar field in the holographic superconductor story~\cite{hartnolletal,Maeda:2009wv}.}  and can be translated into the statement that $A$ and $B$
are analytic functions of the horizon field $\chi_h$. This follows from the fact that $\chi$ is regular at the horizon and the black hole geometry is smooth; we will find a rather different situation at $T = 0$ in the next section.

We now sketch the relevant arguments for finite-$T$. As argued above, for small $\chi_h$ we can expand
$A$ as
 \be \label{anex}
 A \approx a_1 (T) \chi_h + a_2 (T) \chi_h^3  + \cdots
 \ee
and similarly for $B$. Note that both $A$ and $B$ must both be odd functions of $\chi_h$, due to the $\chi \to -\chi$ symmetry.
At $T_c$, a new zero of $A$ should be generated at $\chi_h=0$ and move to a finite value of $\chi_h$ as $T$ is further lowered.\footnote{In fact given $\chi \to - \chi$ symmetry, a pair of zeros are generated and move to opposite directions.} Thus near $T_c$ we
should have $a_1 (T) = a (T - T_c)$ and $a_2 (T) = b + \cdots$ with $a$ and $b$ having the same sign for $T>T_c$. For $T< T_c$, then $A$ has a zero for $\chi_h$ taking a value $\chi_A$ given by
 \be
  \chi_A  = \sqrt{-{a_1 (T) \ov a_2 (T)}} \propto (T_c-T)^\ha  \to 0.
 \ee
Using $\chi_A$ as the initial value at the horizon gives the solution for the condensed phase. Assuming that $B$ is still linear in $\chi_h$ near $\chi_h =0$, we thus find that
 \be
 B (\chi_A) \propto \chi_A \propto (T_c-T)^\ha
 \ee
which gives us the critical exponent $\beta = \ha$. Turning now to the susceptibility, we find
\be
\frac{d B}{d A}\bigg|_{A=0} = \frac{dB}{d\chi_h}\frac{d\chi_h}{d A}\bigg|_{A=0}  \sim (T-T_c)^{-1},
\ee
 leading to $\ga =1$. Furthermore, precisely at $T=T_c$,  $A \sim a \chi_h^3$
and $B$ stays linear near $\chi_h =0$. Thus we find that at $T=T_c$,
\be
B(\chi_h) \propto \chi_h \propto A^{1 \ov 3}
\ee
which gives the critical exponent $\delta = 3$. One can also compute the free energy
using holographic renormalization and indeed find mean field exponent
$F \propto - (T_c - T)^2$ for $T < T_c$. In particular, the analytic expansion~\eqref{anex} for $A$ and $B$ in terms of $\chi_h$ guarantees that $F$ has the Landau-Ginsburg form.

Note that the gravity analysis can be extended to all spacetime dimensions with the same mean field scalings. Presumably this has to do with the fact we are working in the large $N$ limit, which suppresses fluctuations.
This is also consistent with picture obtained in~\cite{Faulkner09} in which
the scalar instability essentially follows from a RPA type analysis. We expect that a $1/N$ computation involving quantum corrections in the bulk  will reveal corrections to these mean-field exponents.

\section{Backreaction and the zero temperature limit}\label{sec:back}

In the usual studies of holographic superconductors, the scalar is charged under a $U(1)$ and so the usual probe approximation involves treating the scalar and $U(1)$ gauge field as negligible perturbations to the background metric. In these models lowering the temperature essentially means increasing the ratio $\mu/T$; thus at sufficiently low temperatures the gauge field (and thus also the condensed scalar) will necessarily backreact strongly on the geometry, causing a breakdown of the probe approximation. The zero temperature limit of the backreacted geometries have been constructed \cite{Gubser:2009cg,Horowitz:2009ij, Gauntlett:2009dn, Gauntlett:2009bh} and typically depend on the details of the couplings and charge of the scalar. One recurring theme is that at zero temperature the charges that was previously carried by the black hole is now completely sucked out of the hole and into its scalar hair. Once the black hole horizon is relieved of the burden of carrying a large charge, it is usually replaced by a degenerate horizon with vanishing entropy.

In our model the situation is different. We always include the backreaction of the gauge field on the metric, as we start from the beginning with the charged black hole solution; it is consistent to solve for a scalar profile on this fixed background only because the scalar is uncharged and so its contribution to the backreaction can be cleanly suppressed by taking $\lambda$ large.

In this section we will discuss the backreacted solution in the IR at zero temperature and argue that even if backreaction is included its effects are rather benign and do not change any qualitative conclusions. This is essentially because all of the charge must stay in the black hole itself, greatly constraining its near-horizon form.

First, we note that Gauss's law states that
\be
\partial_r\left(\sqrt{-g}g^{rr}g^{tt} \partial_r A_t\right) = 0 \; \to \; g_{xx}\sqrt{g^{rr}g^{tt}}\partial_r A_t = \mathrm{const} \ .
\ee
Thus if the electric field is nonsingular at the horizon, $g_{xx}$ must be finite there: this is simply saying that the $\mathbb{R}^2$ at the horizon cannot degenerate as it has a nonzero electric field flux through it. We then expect that the near-horizon geometry factorizes into the form $\sM_2 \times \mathbb{R}^2$, where $\sM_2$ is some 2d manifold involving $(t,r)$. Now consider the trace of the Einstein equation arising from the variation of~\eqref{grac} plus~\eqref{newp}; the $U(1)$ field strength does not contribute to this equation as it is classically scale invariant, and we find:
\be
\mathcal{R} + \frac{12}{R^2} = \frac{1}{2\lambda}\left[(\nabla\chi)^2 + 4V(\chi)\right],
\ee
where $\mathcal{R}$ is the Ricci scalar of the geometry and comes purely from the $\sM_2$ factor. Now let us further assume that in the near-horizon region $\chi$ asymptotes to some constant value $\chi_h$~(we will show this to be consistent shortly). We then find that $\sM_2$ has constant negative curvature, and thus must be $AdS_2$\footnote{More precisely, it could also be an AdS$_2$ black hole; the true zero temperature solution corresponds however to pure AdS$_2$.} with radius $\tilde R_2$ satisfying
\be \label{newr2}
\frac{1}{\tilde R_2^2} = \frac{1}{R_2^2} - \frac{1}{\lambda}V(\chi)
\ee
where $R_2$ is the curvature radius of the original AdS$_2$ near horizon geometry of the
extremal charged black hole. Note that since $V(\chi_h) < 0$, $\tilde R_2 < R_2$.
Thus we see that the backreacted IR geometry is very similar to the unperturbed geometry, except that its $AdS_2$ factor has a radius that is corrected by the presence of the near-horizon scalar potential.

Let us now study the scalar equation of motion \eqref{vve} near the backreacted AdS$_2$ horizon:
\be
\frac{1}{r^2}\partial_r(r^4 \tilde f \partial_r \chi) = R^2 \frac{dV}{d\chi}
\label{scaleqnads2}
\ee
where $\tilde f$ is the warp factor for the backreacted geometry and at zero temperature has a double zero at the horizon $r_0$; expanding near the horizon we see that regularity at the horizon requires
\be
\frac{dV}{d\chi}(\chi(r = r_0)) = 0 \ .
\ee
Thus we see that at the horizon $\chi$ will sit at the bottom of its potential. To understand how this $AdS_2$ region matches onto the asymptotic geometry, we expand $\chi = \chi_h + \delta(r)$ where $\chi_h$ is the bottom of the potential $V(\chi)$. Now working in the AdS$_2$ region and linearizing \eqref{scaleqnads2} near $\chi_h$ we find that $\delta$ obeys the standard $AdS_2$ wave equation:
\be
\partial_r((r-r_0)^2\partial_r \delta) - \tilde R_2^2 V''(\chi_h) \delta = 0,
\ee
whose solutions are
\be
\delta = \al (r-r_0)^{-\frac{1}{2} + \nu} + \beta (r-r_0)^{-\frac{1}{2} - \nu}
\ee
with
\be
\nu = \sqrt{\frac{1}{4} + \tilde R_2^2 V''(\chi_h)},
\ee
Now we would like $\chi$ to approach $\chi_h$ as we approach the horizon; this means that we should not take the $\beta$ solution above, as it invariably blows up as $r \to r_0$. Note that for the $\al$ solution to also not blow up, we need $-\ha + \nu > 0$ and thus $V''(\chi_h) > 0$. To stabilize the AdS$_2$ region we must truly be sitting at a {\it minimum} of the potential at the horizon. In this case the solution
\be
\chi(r) = \chi_h + \alpha(r-r_0)^{-\frac{1}{2} + \nu} \label{nearminsoln}
\ee
can be interpreted as an {\it irrelevant} deformation of the new AdS$_2$ IR CFT that we are flowing to\footnote{Similar considerations are used in \cite{Gubser:2009cg} to determine when an emergent AdS$_4$ can exist.}. The value of the coefficient $\alpha$ is not fixed at this linearized level and must be determined by matching to the UV solution.

We have not solved for the fully backreacted geometry but have used this method to find a normalizable scalar solution on the $T = 0$ charged black hole geometry using the above exponents. Our results match smoothly onto the $T \to 0$ limit of the profiles calculated using the finite temperature matching procedure discussed in Section \ref{sec:AFM}. When $\lam$ is large we do not expect the inclusion of the backreaction
to qualitatively change any of these results given the change of the cosmological constant
is small.

It is instructive to compare this to the situation for charged holographic superconductors \cite{Gubser:2009cg}. In this case for large charge it is found that the IR geometry flows to an $AdS_4$, while for sufficiently small charge it flows to a Lifshitz geometry, with exponent $z$ satisfying
\be
q^2 \sim \frac{1}{z},
\ee
at very small $q$ (see Eq.(81) in \cite{Gubser:2009cg}). We see that our $AdS_2$ solution corresponds to $z = \infty$ at $q = 0$; increasing the charge we have a Lifshitz solution with finite $z$, and finally increasing further we find $z = 1$ for $AdS_4$.

\section{A quantum phase transition from classical gravity} \label{sec:quan}

We recall from the discussion in sec.~\ref{sec:phase} that when $m_c^2 R^2 \equiv -{3 \ov 2}$, the critical temperature approaches zero: thus we should obtain a ``quantum phase transition'' at zero temperature as $m^2$ is varied from above $m_c^2$ to below.
In this section we consider the behavior near the quantum critical
point from the condensed side.

The accessibility of this quantum critical point depends on having the ability to tune the IR dimension of the field $\chi$ in some way. The simplest possible way is by directly tuning the mass, and before proceeding we pause briefly to explain how this may be possible in a simple bulk realization. A similar mechanism is discussed for uncharged black holes in AdS in \cite{Gubser:2005ih}. 

 \subsection{Tuning across the quantum critical point } \label{sec:IVA}

 Imagine that the scalar $\chi$ contains a coupling to another scalar field $\psi$,
\be
S_{\chi} \supset \int d^{d+1} x \, \sqrt{-g} \, F(\psi) \chi^2 \
\ee
where $F(\psi)$ is some function whose detailed form will not be important for us.
Now consider turning on a constant source $h$ for $\psi$ at the boundary, i.e. we require the boundary value of $\psi$ to approach $h$. In the boundary theory language this corresponds to adding a term
\be \label{margD}
\int d^d x \, h \, \sO (x)
 \ee
 to the action with $\sO$ the boundary operator dual to $\psi$. The source induces a nontrivial bulk solution for $\psi$, which in turn contributes to the action of $\chi$ as a mass term, shifting the effective $m^2$. However the new effective mass will typically be radially dependent; we require it to be nonzero at the AdS$_2$ horizon. If we write the IR dimension~\eqref{nuk} of $\psi$ as
\be
\delta_\psi = \ha + \nu_\psi
\ee
then the asymptotics of $\psi$ in the near horizon AdS$_2$ region are
\be
\psi \sim (r-r_0)^{-\frac{1}{2} \pm \nu_{\psi}} \ .
\ee
For $\psi$ to be regular at the horizon we need to choose the $+$ sign in the exponent.
So if $\nu_\psi = \frac{1}{2}$ then $\psi$ approaches a constant at the horizon, and turning on a source $h$ for $\psi$ will change the effective IR mass of the field $\chi$. From~\eqref{nuk}, $\nu_{\psi}$ is $\frac{1}{2}$ if $\psi$ is dual to an operator with UV dimension $\De = d$, i.e. it is massless in the bulk. Thus whenever the UV CFT has such an operator with a nontrivial OPE with the order parameter operator $\Phi$, we expect to be able to tune the system through the quantum critical point by varying $h$ in~\eqref{margD}, i.e. there exists a critical value $h_c$ of $h$ at which we expect a quantum phase transition.
 It would be interesting to understand this mechanism further. Also note that for type II theories in an asymptotic AdS geometry, a natural candidate for $\psi$ is the dilaton.

\subsection{BKT scaling behavior and Efimov states}

For $m^2 < m_c^2$, the BF bound of the near horizon region AdS$_2$ region is violated.
It has been argued in~\cite{Kaplan:2009kr} that for a general AdS gravity dual with mass slightly below the BF bound conformality is lost and an IR scale $\Lambda_{IR}$ is generated exponentially as
 \be \label{IRs}
 \Lam_{IR} \sim \mu \exp \le(- {\pi \ov \sqrt{m_c^2 R_2^2 - m^2 R_2^2}} \ri) \ .
 \ee
where $\mu$ represents some UV scale that in our case is the chemical potential. This exponential behavior is characteristic of the Berezinskii-Kosterlitz-Thouless (BKT) phase transition. One expects that this IR scale controls the physics near the quantum phase transition. In particular, the critical temperature $T_c$ and the value of the condensate $\langle \Phi \rangle$ should be related to this scale.
Here we present a reformulation of the arguments of ~\cite{Kaplan:2009kr} that demonstrates the behavior of $T_c$ and $\vev{\Phi}$ explicitly.

We parameterize the AdS$_2$ region of the uncondensed geometry as
\be
\frac{ds_2^2}{R_2^2} = \frac{-dt^2 + dz^2 }{z^2}.
\ee
The linearized equation for the scalar about the point $\chi = 0$ is then
\be
-\frac{d^2}{dz^2}\chi + \frac{R_2^2(m^2 - m_c^2) - 1/4}{z^2}\chi = \om^2\chi, \label{BFeqn}
\ee
where we have assumed time dependence $e^{-i\om t}$. This is essentially a Schrodinger equation with ``energy'' $\om^2$: the existence of a negative energy bound state indicates a mode growing exponentially in time and hence an instability, and it is a well-known fact that if $z$ can take values from $0$ to $\infty$ then such negative-energy bound states exist when $m^2 < m_c^2$, leading to the BF bound.

In our problem, however, $z$ does not take values on the whole half-line. At some small UV value $z_{UV}$ we should match onto the asymptotic UV geometry. We will now show that if we also choose an appropriate infrared cutoff $z_{IR}$ then even when $m^2$ is below the BF bound \eqref{BFeqn} will have no bound states. To understand this, let us assume for simplicity that $\chi$ satisfies Dirichlet boundary conditions $\chi(z_{IR,UV}) = 0$\footnote{Note that more general boundary conditions simply result in slight changes in the value of $z_{IR}$ and $z_{UV}$} and consider the zero energy $\om = 0$ solutions to \eqref{BFeqn}. We find that the solution is oscillatory (see also discussion in \cite{soojong})
\be
\chi(z) = \sqrt{z}\sin\left[R_2\sqrt{m_c^2 - m^2} \log \frac{z}{z_{UV}}\right], \label{oscsoln}
\ee
and importantly, the zero energy solution satisfies the boundary condition only when
\be
\qquad \log{z_{IR} \ov z_{UV}} = \frac{\pi}{R_2\sqrt{m_c^2 - m^2}} \label{IRUVconst}
\ee
Essentially we must fit a single half-period of the oscillatory wave function inside. This wave function has no nodes; thus if $z_{IR}$ satisfies this condition, then the lowest energy state has zero energy and there is no instability. Decreasing the distance between $z_{IR}$ and $z_{UV}$ will only increase the energy of the ground state. On the other hand, if we {\it increase} this distance--if $z_{IR}$ is too high--then the ground state will have {\it negative} energy, indicating an instability.

\begin{figure}[h]
\begin{center}
\includegraphics[scale=1.0]{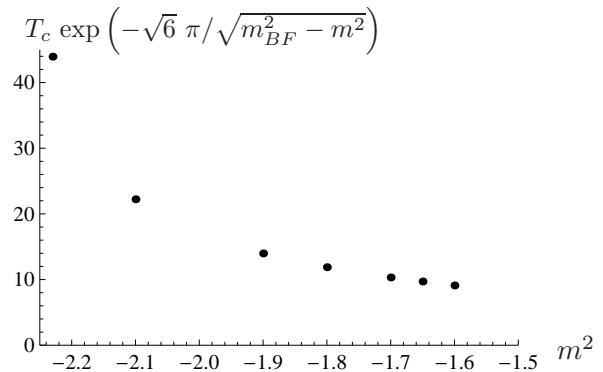}
\end{center}
\vskip -0.5cm
\caption{The exponential dependence of $T_c$ as a function of $\sqrt{m_c^2 - m^2} R_2$. Here we have compensated for the expected theoretical behavior of $T_c$ and plotted the resulting function versus $m^2$; we see that as $m^2 \to m_c^2$ the values approach a constant, verifying our prediction.}
\label{fig:exp}
\end{figure}

Now in our problem $z_{IR}$ can be provided by a small finite temperature, which ends the geometry with a horizon at some large value $z_h$. We thus find that if $z_h > z_{IR}$ as defined above, there will be an instability that can be resolved only by condensation of the scalar. The critical temperature $T_c$ is thus given by
\be
T_c \sim \frac{1}{z_h} \sim \mu \exp\left[-\frac{\pi}{R_2\sqrt{m_c^2 - m^2}}\right]  \label{Tscaling}
\ee
where $\mu \sim \exp[z_{UV}]$ is some UV scale. We have been able to confirm this numerically, including the prefactor in the exponent (see Fig.~\ref{fig:exp}).

What if $T < T_c$? Now the IR cutoff cannot be provided by the horizon, and must instead be provided by nonlinearities associated with the scalar potential, which will become important near $z_{IR}$. We then expect that
the condensate $\langle \Phi \rangle$ should itself exhibit similar exponential scaling
as~\eqref{Tscaling}, as was found explicitly in a similar context in \cite{Jensen:2010ga}. We present here a simple way to understand this result. Note that for $m^2$ only slightly below $m_c^2$ the IR theory is basically the original conformal IR CFT for an exponential hierarchy of scales. In this IR CFT, $\chi$ is unstable and has dimension given by
 \be
 \delta = \ha + i \sqrt{m^2_c R^2_2 - m^2 R_2^2} \approx \ha
 \ee
 Since $\chi$ has dimension $\ha$, it should transform
under a scaling transformation as
 \be
 \chi (\lam z) = \lam^\ha \chi (z) \label{scalchi}
 \ee
We expect that $\langle \Phi \rangle$ should be simply related to the value of $\chi(z)$ at the UV matching value $z_{UV}$, as essentially only linear UV evolution will be required to relate them, and thus
\be
\vev{\Phi} \sim \chi (z_{UV})
\ee
Similarly, the value of $\phi$ in the deep IR will be determined by nonlinearities, which we expect to result in
\be
\chi (z_{IR}) \sim O(1)
\ee
We thus find from \eqref{scalchi} that
 \be
 \chi (z_{IR}) = \sqrt{z_{IR} \ov z_{UV} } \chi (z_{UV})
 \ee
 which results upon evolution through the UV region to the AdS$_4$ boundary in
 \be
\langle \Phi \rangle \sim \sqrt{\frac{z_{UV}}{z_{IR}}} \sim \Lambda_{IR} \exp\left[-\frac{\pi}{2 R_2\sqrt{m_c^2 - m^2}}\right]. \label{phiscal}
\ee
Note that the exponent for $\langle \Phi \rangle$ is half that obtained for $T_c$. In the intermediate conformal regime $\chi$ scales as a dimension $1/2$ operator and the temperature scales as dimension $1$: this is the origin of the factor of two difference in the exponents above. Eventually in the far IR nonlinearities become important and a scale is generated, but the theory is not gapped; provided that the scalar potential has a minimum, as described in Section \ref{sec:back} it instead flows to a different $AdS_2$ fixed point, and the IR scale manifests itself only as an irrelevant perturbation \eqref{nearminsoln} along which we flow to this new fixed point.

\begin{figure}[h]
\begin{center}
\includegraphics[scale=1.2]{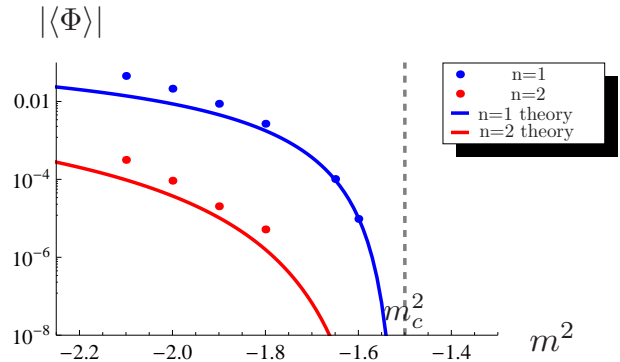}
\end{center}
\vskip -0.5 cm
\caption{Efimov states: for comparison we plot expected results from the analytic arguments described in the text. Agreement can be seen at this level but numerical difficulties prevent us from probing this region more carefully.}
\label{fig:efimov}
\end{figure}

It is possible to perform a more explicit calculation of~\eqref{phiscal}, following the techniques used in~\cite{Jensen:2010ga} where similar scaling is obtained in an AdS$_2$ that arises from a D3/D5 brane construction. From such a treatment it is clear that there should exist an infinite tower of ``Efimov states'' that correspond to allowing oscillatory solutions such as \eqref{oscsoln} to move through more periods before matching to the IR solution. These have $\langle \Phi \rangle \sim \left[-\frac{n\pi}{2 R_2\sqrt{m_c^2 - m^2}}\right]$ with $n$ a positive integer . We have been able to find the first two of these states numerically~(see fig.~\ref{fig:efimov}); as explained in~\cite{Jensen:2010ga} the relevant wavefunctions have more nodes with increasing $n$ and so the $n = 1$ state is energetically favored.

To conclude this subsection we note that turning on any finite temperature the phase transition becomes that of the mean field, since the physics depended in a smooth way on the horizon value of the scalar $\chi_h$. At zero temperature since the horizon is degenerate
we expect the boundary values of $\chi$ to depend on the initial value $\al$ at the horizon in equation~\eqref{nearminsoln}. It would be interesting to understand this better.

\subsection{Quantum critical points for holographic superconductors}

\begin{figure}[h]
\begin{center}
\includegraphics[scale=1]{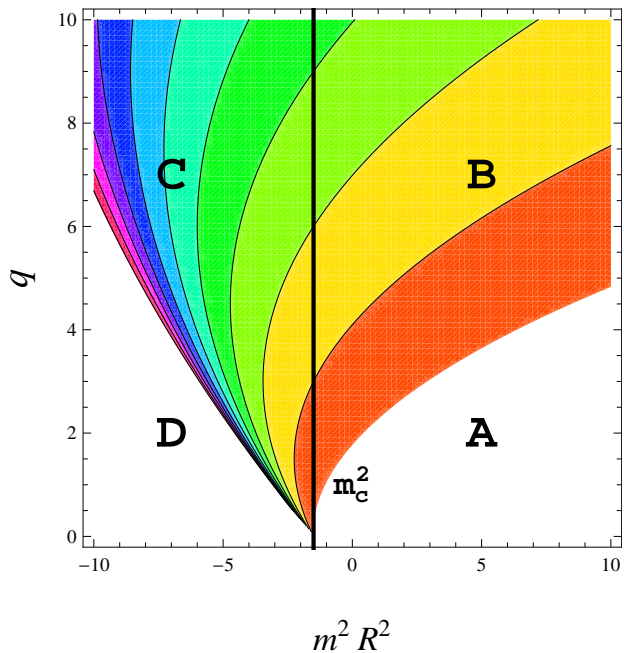}
\end{center}
\vskip -1.0cm
\caption{
Contour plot of the critical magnetic field in the $m^2$, $q$ plane. The color indicates the value of $H_c$, with more purple indicating a higher field. In region $A$ the IR dimension even with $H = 0$ is real and there is no condensate. In regions $B$ and $C$ there exists a finite magnetic field $H_c$ above which the condensate is destroyed. However $H_c$ diverges as we move to the left and is infinite on the boundary between regions $C$ and $D$. In region $D$ even an infinitely strong magnetic field will not stop the condensation. This is understandable: everywhere to the the left of the line $m^2 = m_c^2$ the scalar has a sufficiently negative mass that it would have condensed even if it was uncharged and so it is perhaps not unexpected that a magnetic field cannot halt this condensation. In region $C$ again the scalar would have condensed even if it was uncharged, but here the charge is high enough that a sufficiently strong magnetic field can stop the condensation.}
\label{fig:Bcritical}
\end{figure}

Finally, note that the scaling behavior \eqref{Tscaling} and \eqref{phiscal} also applies to the condensate of a charged scalar, which results in the well-studied holographic superconductor.\footnote{Similar exponential dependence of the critical
 temperature in such a case has been observed by Faulkner and Roberts.} In such a case the coupling to the background electric field is important; also one now has the possibility of supplying an external magnetic field $H$ for $U(1)_{\mathrm{charge}}$\footnote{
 Note that this magnetic field is an external field strength for $U(1)_{\mathrm{charge}}$ and has absolutely nothing to do with the ``magnetic'' field associated with the antiferromagnetic ordering discussed later in this paper, which would correspond to a chemical potential for the $SU(2)$ gauge field $A^{a}_t$.}~\cite{Hartac}. The effect of the electric field on the conformal dimension is shown in~\eqref{nuk}, and the generalization of this formula to include the effect of the magnetic field is (see Appendix~\ref{App:A} for a derivation),
\be \label{begnF}
\delta = \frac{1}{2} + \sqrt{{m^2 R^2 \ov 6} + (6 |bq| -q^2) {\sqrt{1+12b^2}-1 \ov 72 b^2} + {1 \ov 4} }
 \ee
where $b \equiv {H \ov \mu^2_B}$ is dimensionless ratio between the boundary magnetic field $H$ and chemical potential\footnote{$\mu_B$ is the dimensional version of the dimensionless chemical potential $\mu$ we were using earlier, see Appendix~\ref{App:Aa} for further explanation.}. Again the quantum phase transition will happen when $\delta$ becomes complex. It is clear from here that there is a different critical mass $m_c^2$ than in the neutral case; alternatively, one can now imagine tuning the magnetic field through a critical value $H_c$, which is again found by setting the expression inside the square root of~\eqref{begnF} to zero. For $H > H_c$ there will be no condensate, and for $H$ slightly less than $H_c$ one will find similar non-analytic behavior as above as a function of the deviation of $H$ from $H_c$. The explicit expression for $H_c$ is rather complicated and is given in \eqref{bcritical}.

We plot the critical values $H_c$ in the $q-m^2$ plane in Fig.~\ref{fig:Bcritical}. Note in particular that the expression in~\eqref{begnF} containing $b$
saturates at a finite value $|q| \ov {2 \sqrt{3}}$ as $b \to \infty$. Thus if
 \be
 m^2 R^2 + \sqrt{3} |q| < -{3 \ov 2}
 \ee
no matter how large the magnetic field is, a condensate cannot be prevented.
This is region $D$ in Fig.~\ref{fig:Bcritical}. This is understandable: everywhere in this region the scalar mass is below the neutral AdS$_2$ BF bound $m^2R^2 = -\frac{3}{2}$ and so it would have condensed even if it was uncharged. As the electric field is in some sense not responsible for the instability, it may make sense that a magnetic field cannot halt it.
It will be interesting to find field theoretical models with this feature.

\section{Antiferromagnetism and Spin Waves} \label{sec:emb}

We have understood in detail the mechanism by which a neutral scalar field can condense in a finite-density geometry. We would now like to employ this new understanding to holographically model symmetry breaking with such a neutral order parameter. One immediate example is antiferromagnetic order: as explained in the introduction, in the low-energy limit of interest to us spin rotations may be thought of as a global $SU(2)$ symmetry. Antiferromagnetic ordering then corresponds to a spontaneous breaking $SU(2) \to U(1)$, where the unbroken direction corresponds to rotations about a single axis. In such a system the background value of the spin density remains zero, but there is a neutral order parameter $\Phi^a$ that corresponds to the staggered magnetism.

In this section we construct the bulk dual of the operator $\Phi^a$: we embed the neutral scalar field $\chi$ discussed in sec.~\ref{sec:AFM}--\ref{sec:quan} into part of an $SU(2)$ triplet scalar field $\phi^a$ charged under the $SU(2)$ bulk gauge field corresponding to the global spin $SU(2)$ symmetry of a boundary theory. The phase transition discussed earlier then becomes a transition to an antiferromagnetic (AFM) phase.

Note that while we use the word ``antiferromagnetic'' there is no sense in which the microscopic degrees of freedom of our system consist of spins that are anti-aligned on a bipartite lattice. We use the term to describe only the symmetry breaking pattern described above, realized in a manner that manifestly requires a finite density for symmetry breaking. We do feel that if indeed gravity duals could be constructed top-down for such systems they would likely contain ingredients similar to those in our description.
\subsection{Embedding of $\chi$}
More explicitly, we can consider the following action
 \be
 S = S_{\rm grav} + S_{\rm matter}
 \ee
 with
 \be
 S_{\rm grav} = \frac{1}{2\kappa^2} \int d^4 x\sqrt{-g} \left(\mathcal{R} + \frac{6}{R^2} \right)
\ee
where $R$ is the curvature radius of AdS$_4$.
 The relevant part of the bulk Lagrangian for matter fields is then given by
  \bea
 {2 \kappa^2 \ov R^2} \sL_{\rm matter} &=& - {1 \ov 4 g_A^2} F_{MN}^{ a} F^{MN a} -  G_{MN} G^{MN} \cr
 && - \frac{1}{\lambda}\left(\ha (D \phi^a)^2  - V(\phi^a)\right)
 \label{master2}
 \eea
with
\bea
&& F_{MN}^a = \p_M A_N^a - \p_N A_M^a + \ep^{abc} A_M^b A_N^c, \cr
&& G_{MN} = \p_M B_N - \p_N B_M,
 \eea
 and
 \be
 D_M \phi^a = \p_M \phi^a + \ep^{abc} A_M^b \phi^c \ .
  \ee
We will take the potential $V$ to have the double well form of~\eqref{newp}
 \be
 V(\phi^a) = \ha m^2 \vec \phi \cdot \vec \phi + {1 \ov 4} (\vec \phi \cdot \vec \phi)^2
  \ .
 \ee

We again put the system at a finite density by turning on a chemical potential for the $U(1)_{\rm charge}$, with a background metric given by~\eqref{bhmetric1}.
We note that in this background all $SU(2)$ gauge fields are inactive. This highlights an important difference between this setup and the usual $U(1)$ holographic superconductor; in the Abelian case there is a background chemical potential $\mu$ for the $U(1)$ charge; this does not break the $U(1)$ but does interact via the bulk equations of motion with the charged scalar order parameter, causing it to condense for a suitable choice of mass and charge. This is distinctly different from the $SU(2)$ case studied here; specifying a chemical potential for the $SU(2)$ would involve picking a direction $\mu^a$ in the $SU(2)$ space, corresponding to an {\it explicit} breaking of the symmetry. This is analogous to applying an external magnetic field. We study this explicit breaking in sec.~\ref{sec:fmw} and for now focus on the spontaneous breaking of $SU(2)$.

When $m^2$ falls in the range~\eqref{instcrit}, the background \eqref{bhmetric1} becomes unstable towards condensation of the scalar $\phi^a$, which will then
\emph{spontaneously} break the $SU(2)$ to a $U(1)$ subgroup. More explicitly,
consider the following ansatz
\be \label{ansS}
\phi_0^a = \le(0,0, {\chi(r) \ov R}\ri), \qquad A_M^a = 0
\ee
which can be readily checked by examining the equations of motion
as being self-consistent, if we ignore the backreaction to the background geometry.
We note in particular that one can consistently set the $SU(2)$ gauge field to $0$.
This is because of the non-Abelian nature of the interactions of the gauge field and scalar, e.g. $f^{abc}\phi_b A_c$ etc. which clearly vanish if all objects point only in one direction in the $SU(2)$ space.

Plugging~\eqref{ansS} into the equations of motion following from~\eqref{master2} we precisely find~\eqref{vve} and our previous discussion of the phase transition in sec.~\ref{sec:AFM}--\ref{sec:quan} can be taken over completely. Note that if we consider a finite $\lam$ and turn on backreaction, the profile for the $U(1)$ gauge field in~\eqref{bhmetric1} and the background metric will be modified, but one can still set the $SU(2)$ gauge field to zero.

\subsection{Spin waves}
To characterize the ordered phase, we study in this section
perturbations around the symmetry breaking solution~\eqref{ansS}  with a nonzero order parameter $\phi^a$ but zero background gauge field $A^a_M$. We show that the system has two linearly dispersing gapless modes, corresponding in the field theory to the two Goldstone modes arising from the spontaneous breaking of the global $SU(2)$ symmetry, and that their velocity is given by the
expression expected
from the standard field theory for a quantum antiferromagnet.
 From the bulk point of view, low-frequency rotations of the order parameter in the broken symmetry directions will source bulk gauge field fluctuations, which we will find to be normalizable at the $AdS$ boundary if they obey a specific dispersion\footnote{Methods similar to those used here may be used to obtain analytic control over the hydrodynamics of the U(1) holographic superconductor \cite{nabilseandio}.}.

Our analysis for the remainder of this section will not depend at all on the details of the metric or scalar profile (except that it is normalizable). We will work in a general boundary spacetime with a bulk black brane metric given by
\be \label{speMe}
ds^2 \equiv g_{MN} dx^M dx^N = g_{tt} dt^2 + g_{rr} dr^2 + g_{xx} d \vec x^2
\ee
We work at finite temperature with $g_{tt}$ ($g_{rr}$) having a first order zero (pole) at the horizon $r = r_0$.

Let us begin by understanding \emph{why} from the gravity point of view there should exist a gapless mode. Consider a global $SU(2)$ gauge rotation of the background order parameter $\phi_0^a (r)$, at $\om = 0$ and $k = 0$.
\be
\delta \phi^a_i (r) = \epsilon \tau^i_{ab}\phi_0^b(r) \label{scalgauge}
\ee
where $\tau^i, i=1,2$ is a generator along one of the broken directions so that $\delta\phi^a$ is nonzero and $\ep $ is a constant. This small perturbation will obviously be a solution to the bulk gravity equations of motion; it is in fact a normalizable solution at the AdS boundary, precisely because $\phi_0^a$ is normalizable. This appears somewhat trivial---but note that this is exactly a gapless mode, as it is a normalizable solution that exists in the limit $\om \to 0$. Now consider a {\it local} $SU(2)$ gauge rotation with
the gauge parameter $\ep$ having a small frequency and momentum in the $x$-direction, i.e. $\ep (t,x) \propto e^{- i \om t + i k x}$. The perturbation to the scalar takes the same form \eqref{scalgauge}, but now we have extra perturbations to the gauge fields
\be
\delta A_t^i = -i\om\epsilon \qquad \delta A_x^i = ik\epsilon \ .
\ee
These are \emph{not} normalizable at infinity, and thus this pure gauge transform is no longer a normalizable solution. Thus at any nonzero $\omega$ and $k$ to find a normalizable mode we must move off of the pure gauge solution and actually solve the dynamical equations of motion. The existence of the the global solution \eqref{scalgauge} guarantees that we as we take $\om, k \to 0$ we will find a gapless solution, but the dynamical bulk equations of motion will show that it will be normalizable only when a certain dispersion relation $\om(k)$ is satisfied.

\subsubsection{Pions in the bulk} \label{moti}
The previous discussion has convinced us that the existence of the gapless mode is intimately related to our ability to perform global rotations on the order parameter, i.e. to \emph{bulk} Goldstone modes. Let us thus parametrize fluctuations of $\phi$ in terms of bulk Goldstone fields (or ``pions'') $\pi^i(r,t,x)$ as follows:
\be
\phi(r,x) = \exp(i\pi^i(r,t,x)\tau^i)\phi_0(r)
\ee
where $\phi_0(r)$ is the background solution and as before $\tau^i$ are the broken symmetry generators. We now work out the bulk quadratic action of the $\pi^i$ to be\footnote{ In writing this expression we have assumed that the background gauge field is zero; thus we have neglected intrinsically non-Abelian terms of the form $f^{abc}\pi_b A_c$ which all arise at higher order.}
\be
S(\pi) = -\frac{R^2}{2\kappa^2}\int d^4 x \frac{1}{2\lambda}\sqrt{-g} g^{MN}(\partial_M \pi^i - A_M^i)(\partial_N \pi^j - A_M^j)h_{ij}
\ee
Here $h_{ij}$ can be viewed as a metric on the Goldstone boson space, given by
\be
  h_{ij}(r) = \ha \phi_0^{\dagger}(r)\{\tau_i, \tau_j\}\phi_0(r)
  = \frac{\chi^2}{R^2} \delta_{ij}
\ee
Thus the two Goldstone modes decouple. Below we will focus on one of them and drop
the index $i$ for notational simplicity.

The equation of motion for $\pi$ is then
  \be \label{eom1}
\partial_M\left(\chi^2 \sqrt{-g}g^{MN}(\partial_N\pi - A_N)\right) = 0
\ee
It is clear that in the limit $\om, k \to 0$, a constant pion profile $\pi(r) = \pi_0$ and vanishing gauge field $A = 0$ provide a solution to the equations of motion. This is the gauge transform discussed earlier.

Let us now briefly detour to discuss the asymptotic behavior of the pion field. Carrying out the standard analysis and using the fact that the background solution is normalizable\footnote{We restrict ourselves to the ordinary quantization for this section.}: $\chi (r) \sim r^{-\Delta}$ for large $r$, we find that as $r \to \infty$.
\be
\pi(r) \approx B + A r^{-d+2\Delta}, \qquad r \to \infty \label{piasymp}
\ee
To understand this it is useful to remember that fluctuations in the original field $\delta\phi(r)$ are related to $\pi$ as $\delta\phi(r) = \pi(r)^i \tau^i \phi_0(r) \sim \pi(r)r^{-\Delta}$. Comparing this to the asymptotic behavior of the scalar~\eqref{scalasymp} we see that the \emph{constant} piece $B$ of $\pi$ at infinity is actually \emph{normalizable}, while the piece $A$ is not normalizable and should be viewed as the source for the pion field.\footnote{It is amusing to note that computing the momentum conjugate to $\pi$ precisely extracts out the \emph{source} in this case, in exactly the same way that it extracts out the vev in the more familiar case of a standard massless scalar.}

\subsubsection{Yang-Mills equations}

Fluctuations of $\pi$ will excite the non-Abelian gauge field $A^a_M (r,t,x)$, whose linearized
equations are
\be \label{maxw}
\frac{1}{\tilde{g}_A^2} \nabla_M F^{MN} + \chi^2 g^{NP}(\partial_P\pi - A_P) = 0 \ .
\ee
Again we have suppressed the index $i$ which labels the broken
generators, since at the linearized level different directions decouple. We have absorbed various factors into a rescaled gauge coupling
\be
\frac{1}{\tg_A^2} = \frac{R^2\lambda}{g_A^2},
\ee
which controls the strength of the effects of the scalar sector on the gauge field. We could of course always choose a unitary gauge to get rid of the Goldstone boson $\pi$---but
to make the connection to boundary Goldstone modes more obvious we will not do this and instead choose the gauge $A_r =0$. It is convenient as in~\cite{Iqbal:2008by} to work with the momenta $J^{\mu}$ conjugate to the bulk gauge field $A_{\mu}$, defined as\footnote{For convenience of discussion here we have chosen a different normalization for the currents below.}
 \be
J^\mu \equiv \frac{1}{\tg_A^2} \sqrt{-g}F^{\mu r} = -\frac{1}{\tg_A^2} \sqrt{-g} g^{rr} g^{\mu \nu}
\p_r A_\nu \ .
\label{jdef}
\ee
The value of the bulk field $J^{\mu}$ at the AdS boundary $r \to \infty$ is equal to the expectation value of the field theory current $\langle \sJ^{\mu} \rangle_{\mathrm{QFT}}$.

The $N = r$ component of the Maxwell equations~\eqref{maxw} can now be written
\be
\partial_{\mu} J^{\mu} = - \sqrt{-g}g^{rr} \chi^2(r)\partial_r\pi \ .
\label{consj}
\ee
If the symmetry is unbroken then $\chi = 0$ and this is nothing but the conservation of current. Let us now consider evaluating this expression at the AdS boundary. Comparing~\eqref{consj} to the asymptotic behavior of $\pi$ in \eqref{piasymp} we see that the entire right-hand side of this expression is proportional (with no extra factors of $r$) to the coefficient $A$ in the near-boundary expansion of $\pi$, i.e. to the source for the Goldstone boson field. This is simply the usual field theory Ward identity, which says that in the absence of a Goldstone source the field theory current is conserved.

The other nontrivial equations from~\eqref{maxw} are those in $t$ and $x$ directions with nonzero $A_x (r,t,x)$ and $A_t (r,t,x)$ only, given by
\begin{align}
\label{eom2}
-\p_r J^t + {1 \ov \tg_A^2} \nab_x F^{xt}+ \sqrt{-g} g^{tt} \chi^2 (\p_t \pi - A_t) & = 0 \\
-\p_r J^x + {1 \ov \tg_A^2} \nab_t F^{tx} + \sqrt{-g} g^{xx} \chi^2 (\p_x \pi - A_x) & = 0
 \ .
\label{eom3}
\end{align}
Equations~\eqref{eom1} and~\eqref{consj}--\eqref{eom3} are the full set of equations for this system.

\subsubsection{Boundary Goldstone modes and their spin wave velocity}
We will now solve the above equations in the hydrodynamic limit to show explicitly the existence of a Goldstone mode, i.e. a normalizable solution to ~\eqref{eom1} and~\eqref{consj}--\eqref{eom3} that is infalling (or regular) at the horizon and exists in the limit of small $\om$ or $k$. We work in Fourier space
\be
\pi = \pi(r)e^{-i \om t + i k x} \qquad A_{t,x} = A_{t,x}(r) e^{-i\om t + i k x}
\ee
Recall that at precisely $0$ frequency and momentum we already know a solution to these equations: a constant pion profile $\pi(r) = \pi_0$ and $A = 0$. We now expand around this solution in powers of $\om$ and $k$. To lowest order we simply solve the {\it forced} equations for $A$ given by \eqref{eom2} and \eqref{eom3} with the forcing term provided by the zeroth-order solution for $\pi$, meaning that
\be
A_t(r) \sim \sO(\om\pi_0), \qquad A_x(r) \sim \sO(\pi_0k).
\ee
The first correction to $\pi(r)$ enters through \eqref{eom1} and involves one extra field theory derivative, and thus we see that
\be
\pi(r) = \pi_0 + \pi_1(r) \qquad \pi_1(r) \sim \sO(\pi_0\om^2, \pi_0k^2)
\ee
The solution we want for $A$ is both infalling at the black hole horizon and normalizable at the boundary; such a solution is nontrivial because of the forcing term\footnote{Of course as the force $\pi_0$ vanishes the only solution that is infalling and normalizable becomes the trivial one $A = 0$.} and is given by:
\be \label{newd}
A_t = - i \om \pi_0 \left(1 - a_t(r)\right) , \quad A_x =  i k \pi_0 \left(1 - a_x(r)\right)
\ee
where $a_t$ and $a_x$ are defined to be infalling at the horizon and satisfy the homogenous part (and zero frequency) part of \eqref{eom2} and \eqref{eom3}:
\begin{align}\label{Aeqns}
\frac{1}{\tg_A^2} \partial_r(\sqrt{-g}g^{rr}g^{tt}\partial_r a_t) - \sqrt{-g} g^{tt} \chi^2 a_t & = 0  \\
\frac{1}{\tg_A^2} \partial_r(\sqrt{-g}g^{rr}g^{xx}\partial_r a_x) - \sqrt{-g}
g^{xx} \chi^2  a_x & = 0
\label{aeq2}
\end{align}
To ensure that $A_t$ and $A_x$ in~\eqref{newd} are normalizable we also require that
\be \label{erp}
a_t (\infty) = 1, \qquad a_x (\infty) =1 \ .
\ee

The corresponding equations for the pion profile are
\be
\pi_1 = \om^2 \pi_0 C^t(r) + k^2 \pi_0 C^x(r),
\ee
where the dynamical equations for $C^t$ and $C^x$ can be written from \eqref{eom1} as expressions for the radial independence of two objects $\alpha^t$ and $\alpha^x$:
\begin{align}
\label{x1}
\partial_r \alpha^x = 0, \; \al^x & \equiv - \sqrt{-g} \chi^2 g^{rr} \partial_r C^x + \frac{1}{\tg_A^2} \sqrt{-g}g^{rr}g^{xx}\partial_r a_x \\ 
\label{x2}
\partial_r \alpha^t = 0, \; \al^t & \equiv \sqrt{-g} \chi^2 g^{rr} \partial_r C^t - \frac{1}{\tg_A^2} \sqrt{-g}g^{rr}g^{tt}\partial_r a_t 
\end{align}
The precise form of these equations turns out to not be important: the crucial fact is that $C^{t,x}$ are essentially driven by $a_{t,x}$, and so we can find solutions to them that are also infalling and normalizable--where normalizable in this case means that the term $\chi^2\partial_r C^{x,t}$ vanishes at infinity (as explained in the discussion around \eqref{piasymp}).

So far it appears that we have found a solution that is both infalling and normalizable, for {\it all} small frequency and momenta. This should not be possible, and indeed we have yet to impose the constraint \eqref{consj}.  We find then the relation
\be \label{souv}
\om^2 = v_s^2 k^2 \qquad v_s^2 = \frac{\alpha_x}{\alpha_t}\ .
\ee
This is consistent only because of the radial independence of $\alpha_x$ and $\alpha_t$, which away from the boundary involves fluctuations of the pion field.
Thus we have shown that there exists a normalizable infalling mode provided that $\om,k$ obey a linear relation. This is our main result.

It is convenient to evaluate~\eqref{x1} and~\eqref{x2} at $r =\infty$, where
the terms depending on $C^{t,x}$ do not contribute, which leads to
 \begin{align}
\alpha_x &= \frac{1}{\tg_A^2}\lim_{r \to \infty} \sqrt{-g}g^{rr}g^{xx}\partial_r a_x , \\ \alpha_t & = -\frac{1}{\tg_A^2}\lim_{r \to \infty} \sqrt{-g}g^{rr}g^{tt}\partial_r a_t \ .
\end{align}
Note that this is equivalent to the statement that at the boundary the right-hand side of \eqref{consj} vanishes---in the absence of a Goldstone boson source the current is conserved.
Recall that $a_t, a_x$ obey the zero frequency and unitary gauge infalling wave equations. It was shown in \cite{Iqbal:2008by} that on such a field configuration at infinity the ratio of the current $J_{x,t} \sim \partial_r a_{x,t}$ to the field $a_{x,t}$ (which are unity in this case from~\eqref{erp}) itself is simply the field theory Green's function for the current. We thus conclude that
 \begin{align}
 \al_x = \lam R^2 G_{xx} (\om = 0, k = 0), \\
  \al_t = - \lam R^2 G_{tt}(\om = 0, k = 0)
 \end{align}
 where
 \be
 G_{\mu \nu} (\om, k) = \vev{j_\mu (\om,k) j_{\nu} (-\om,k)}_{\rm retarded}
 \ee
 are momentum space retarded Green function of the spin current $j_{\mu}$
along a symmetry broken direction. We thus find that the spin velocity
 $v_s$ can be written as
  \be \label{ope}
  v_s =  \left(\frac{G_{xx}(\om = 0, k = 0)}{-G_{tt}(\om = 0, k = 0)}\right)^{1/2} \ .
  \ee
This is the expected expression for an antiferromagnet.
In particular,
recognizing that \cite{Haldane83,Chakravarty89,Auerbach_book,Chandra90}
 \be
 \rho_s = G_{xx}(\om = 0, k = 0), \qquad \chi_\perp = -G_{tt}(\om = 0, k = 0)
 \ee
 where $\rho_s$ is the spin stiffness and $\chi_\perp$ transverse magnetic
 susceptibility, we recover the standard expression
  \be
  v_s = \sqrt{\rho_s \ov \chi_\perp }  \ .
  \ee

Note that equation~\eqref{newd} has a simple boundary theory
interpretation: $j_\mu \propto \p_\mu \pi_0$, as expected for the
superfluid part of the current density. In our probe analysis, there is no normal component. Also note that our analysis above gives a nice correspondence between the Higgs mechanism in the bulk and the dynamics of Goldstone modes in the boundary theory.

\subsection{Evaluation of spin wave velocity}
We now evaluate $G_{xx}$ and $G_{tt}$ to find the spin velocity. It is convenient to derive a flow equation for them
as in~\cite{Iqbal:2008by}. We start with $a_x$; examining the near-horizon behavior of equation~\eqref{Aeqns} we see that for the solution to be nonsingular we need
\be
\frac{\partial_r a_x}{a_x} = \frac{\tg_A^2 \chi^2}{4\pi T} \label{Gxbc}
\ee
where $T$ is the temperature. Introducing
\be \label{g1}
\sG_{x}(r) =  {1 \ov \tg_A^2} \frac{\sqrt{-g}g^{rr}g^{xx}\partial_r a_x}{a_x}
\ee
from~\eqref{aeq2} we find $\sG_{x}$ satisfies the flow equation
\be
\partial_r \sG_{x}(r) = \sqrt{-g}g^{xx}\chi^2  - \frac{\tg_A^2 \sG_{x}^2 (r)}{ \sqrt{-g}g^{rr}g^{xx}} \label{Gxflow}
\ee
This equation should be integrated from the horizon (where from \eqref{Gxbc} we have $\sG_x(r_h) = 0$) to the AdS boundary, where it is equal to the Green's function $G_{xx}(\om=0, k = 0)$.

\begin{figure}[h]
\begin{center}
\includegraphics[scale=0.30]{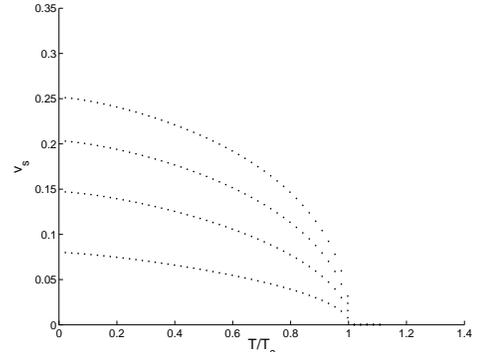}
\end{center}
\vskip -1.0cm
\caption{Spin wave velocity as a function of $T/T_c$ for various values of the rescaled gauge coupling $\tg_A$: $\tg_A$ is varied from $1$ (lowest curve) in unit increments to $4$ (highest curve).}
\label{fig:svel}
\end{figure}

Similarly to above, for $a_t$ we introduce
\be \label{h1}
\sH_t(r) = \frac{ \tg_A^2 a_t}{\sqrt{-g}g^{rr}g^{tt}\partial_r a_t}
\ee
This is convenient, as at the horizon $a_t =0$. The corresponding flow equation for $\sH_t$ is
\be
\partial_r \sH_t(r) = \frac{\tg_A^2}{\sqrt{-g}g^{rr}g^{tt}} - \sH_t^2 (r) \sqrt{-g}g^{tt}\chi^2
\ee
and it should be integrated from the horizon, where the relevant boundary condition is $\sH_t(r_h) = 0$, to infinity. In terms of~\eqref{g1} and~\eqref{h1} the spin wave velocity~\eqref{ope} can be written as
\be
v_s = \left(-\sG_x(\infty) \sH_t(\infty)\right)^{\frac{1}{2}} \ .
\ee
Evaluation of this requires knowledge of the scalar profile and can only be done numerically; some representative plots are shown in fig.~\ref{fig:svel}.

It is interesting to see what happens near the phase transition, where we have some analytic control; here we have $\chi^2 \sim (T-T_c)$ from earlier results; now examining the flow equations we see that $\sH_t(\infty)$ remains finite in this limit, and is in fact given (up to corrections analytic in $(T-T_c)$) by
 \be \label{soem}
 -\sH_t = - \int_{\rm r_0}^\infty \frac{\tg_A^2}{\sqrt{-g}g^{rr}g^{tt}}  = {1 \ov \Xi}
 \ee
 where $\Xi$ is the spin susceptibility in the unbroken phase. The second equality in~\eqref{soem} follows
 from the $U(1)$ analysis in Appendix D of~\cite{Iqbal:2008by}. On the other hand for $\sG_x$ we find from~\eqref{Gxflow} that near the transition $\sG_x \sim (T-T_c)$. Thus the nonlinear term in the evolution can be neglected, leading to
 \be
 \sG_x (\infty) = \int_{r_0}^\infty dr \, \sqrt{-g}g^{xx}\chi^2 \ .
 \ee
We find then near $T_c$
  \be
  v_s^2 = {1 \ov \Xi} \int_{r_0}^\infty dr \, \sqrt{-g}g^{xx}\chi^2 \propto (T- T_c)\ .
  \ee
Note that above $T_c$, $\rho_s = G_{xx} (\om =0, k=0)$ becomes identically zero. It would be interesting to extend this analysis to understand what happens to the spin wave velocity at zero temperature near the quantum critical transition.

\section{External ``magnetic fields'' and forced ferromagnetic magnons} \label{sec:fmw}

We have constructed a gravity description of the spontaneous breaking of an $SU(2)$ symmetry--analogous to the Neel phase in a spin system--and displayed the associated spin waves. In this section we turn briefly to a ferromagnet. A ferromagnet can be understood as a system undergoing spontaneous symmetry breaking in which the broken vacuum is charged under the unbroken generator.
The spin waves in this case possess
a quadratic dispersion $\om \sim k^2$.

This setup is not straightforward to realize in holography--the spontaneous generation of a nonzero spin density means that essentially one now wants the {\it unbroken} non-Abelian gauge field {\it itself} to condense, without supplying any external chemical potentials. We leave this for future work.

In this section we consider something simpler--imagine applying an external magnetic field $H$ to a sample containing spins, forcing them to align along the direction of the field. In our setup this corresponds to picking a direction in the $SU(2)$ space -- we will pick the $3$ direction -- and supplying a chemical potential $\mu \sim H$ for the gauge field in that direction (in this section only $\mu$ refers to the chemical potential for the non-Abelian gauge field; we will set the $U(1)$ chemical potential to $0$ here, as it does not play an essential role). It is interesting to note that this model has been studied before: it is precisely the ``normal phase'' of the holographic $p$-wave superconductor studied by \cite{Gubser:2008wv}, but our interpretation is different: in particular in those models a $U(1)$ subgroup of the $SU(2)$ is usually taken to be electric charge, which is quite different from our approach. We will show that small fluctuations around such a configuration exhibit spin waves whose dispersion is not gapless, but whose dependence on momentum is indeed quadratic.

\subsection{Gravity setup}

We consider a generic spacetime metric~\eqref{speMe} and a nontrivial non-Abelian gauge field profile
\be
A_t^3(r) = \mu\alpha(r) \qquad \alpha(r \to \infty) = 1 \qquad \alpha(r_0) = 0, \label{bgfmag}
\ee
with all other matter fields turned off. In particular the scalar $\phi^a$ is uncondensed and will play no role. We will also set the gauge coupling $g_A$ to $1$ as when the scalar is inactive $g_A$ plays no role in the classical dynamics that follows. The profile $\al (r)$ solves the equation
 \bea
&& {1 \ov \sqrt{-g}} \p_r \le(\sqrt{-g} g^{rr} g^{tt} \p_r \al (r) \ri) = 0 \\
&&\Rightarrow
 \quad J^{3t} \equiv - \sqrt{-g} g^{rr} g^{tt} \p_r A_t^3 = {\rm const}
 \eea
where as above $J^{3t}$ is the canonical momentum conjugate to $A_t^3$ and gives the expectation value of the spin density $j^{3t}$ in the boundary theory as $\langle j^{3t}\rangle = J^{3t}(r = \infty)$. For simplicity, we will ignore the backreaction to the background metric, requiring that both $\mu$ and $J^{3t}$ are small when expressed in units of the temperature. Note that $\mu$ and $\sJ^{3t}$ are related by the spin susceptibility $\Xi$
 \be \label{susC}
\langle j^{3t} \rangle = \mu \Xi, \qquad
 \ee
with $\Xi$ given by~\eqref{soem}.

We will now slightly perturb this solution. We do not expect to find a gapless mode, as we have an ``external magnetic field''; however at $k = 0$ we do expect to find a normalizable mode with $\om = \mu$. To see the existence of this mode consider performing a gauge transform on the background \eqref{bgfmag} with infinitesimal gauge parameter $\Lambda^a$:
\be
\delta A^{a}_M = \ep^{abc}\Lambda^c A_M^b + \partial_{M}\Lambda^a
\ee
It is convenient to work with the linear combinations
\be
A^{\pm} = A^1 \pm i A^2,
\ee
as these are charge eigenstates under the rotations in the $3$ direction for which we have supplied a chemical potential. If we assume a time dependence of the form $e^{-i\om t}$ for $\Lambda^+$ we find the resulting $A_t^+$ can be written as
\be
 A^{+}_t = \Lambda ( A^3_t - \om ) e^{- i \om t}
\ee
where $\Lambda$ is an overall (complex) constant. This perturbation is generically not normalizable at the AdS boundary; however if we demand that $ A^{+}_t$ vanish at $r \to \infty$ then this fixes the frequency and we find
\be
\om = \mu,  \qquad  A^{+}_t(t,r) = \Lambda \mu(\alpha(r) - 1) e^{- i \mu t}
\equiv \Lambda A_{t0}(r)  e^{- i \mu t}
 \label{nAbgauge}
\ee
Thus there is a normalizable mode at $\om = \mu$ with the specified radial profile $A_{t0}(r) = \mu(\alpha(r) - 1)$\footnote{The existence of such a mode was observed before in~\cite{Gubser:2008wv}
 in the context of conductivity of a p-wave superconductor.}.
We will now turn on a small $k$ to see how this mode evolves; as expected for a system with a background spin density, we will find that the frequency has a quadratic dependence on $k$.

\subsection{Dispersion relation for the ferromagnetic magnon}

To do this, we will need the full bulk Yang-Mills equations:
\be \label{w3}
 {1 \ov \sqrt{-g}} \p_M \le(\sqrt{-g} F^{aMN} \ri) + \ep^{abc}A^b_M F^{cMN} = 0\ . 
\ee
The presence of the background gauge potential $A^3_t$ can be conveniently taken into account by defining a gauge-covariant partial time derivative $d_t$ that acts in the $\pm$ basis as
\be
d_t A^{+}_{t,x} = (\partial_t + i A_t^3) A^{+}_{t,x} \ .
\ee
We give all fields a spacetime dependence $e^{-i\mu t}e^{-i\Om t + i kx}$. Here $\Om \equiv \om - \mu$ will parametrize deviation from the $\om = \mu$ solution found above. The relevant equations of motion are those in \eqref{w3} for $N=(r,t,x)$. We will work again with the bulk canonical momenta, defined as before:
\bea
J^{+t} &\equiv & -\sqrt{-g} F^{+rt} = -\sqrt{-g} g^{tt} g^{rr} \p_r A_t^+, \\
 J^{+x} &\equiv& -\sqrt{-g} F^{+rx} = -\sqrt{-g}  g^{xx} g^{rr} \p_r  A_x^+ \ .
\eea
We first examine the $N = r$ component of \eqref{w3}, which can be written as
\be
\label{nonAbcons}
d_t J^{+t} + \partial_x J^{+x} = i J^{3t}A_t^{+} .
\ee
This equation is a bulk constraint that reduces on the boundary to the non-Abelian conservation of current. The remaining dynamical equations are
\bea
 \label{w4}
&&  {1 \ov \sqrt{-g}} \p_r J^{+t} + \p_x F^{+xt} =0 \\
&&  {1 \ov \sqrt{-g}} \p_r J^{+x} + d_t F^{+tx} =0 \
\label{w5}
\eea
Only one of the components of the field strength tensor is affected by the gauge field background:
\be
F_{xt}^+ = \partial_x A_t^{+} - d_t A_x^{+} = i k A_t^+ - (i \mu (\al -1) - i \Om) A_x^+
 \ .
\ee
We now want to perturb around the $\om = \mu$ solution found above. We thus turn on a small $k$-dependence in~\eqref{nAbgauge} and consider a perturbation of the form
 \be \label{w1}
  A_t^+ = e^{- i \mu t} e^{- i \Om t + i k x} \le( A_{t0} (r) + A_{t1} (r; w,k) + \cdots \ri)
   \ee
where $A_{t0} (r) \sim (\alpha(r) - 1)$ is the previously found profile in~\eqref{nAbgauge}. Recall that $\Om \equiv \om - \mu$ parametrizes departure from $\mu$ and will be the small parameter in the expansions that follow. A nonzero $A_t^+$ will also excite an $A_x^+$ which we take to have the form
\be \label{w2}
A_x^+ = A_x^+(r;\Om,k) e^{- i \mu t} e^{- i \Om t + i k x} \ .
\ee
The other components of the gauge fields can be consistently set to zero except for $A^{-}_{t,x}$, which are related to $A^{+}_{t,x}$ by complex conjugation.

To obtain the magnon dispersion relation we will expand the above equations to lowest order in the $\Om$ and $k$ expansion and look for solutions that are both infalling (or regular) at the horizon and normalizable at the AdS boundary. These equations will have the same structure as in the antiferromagnetic case above; we will find second-order radial equations for $A_x^+$ and the correction $A_{t1}$ that are forced by the known solution $A_{t0}$.

More explicitly, plugging~\eqref{w1} and~\eqref{w2} into~\eqref{w4}--\eqref{w5}, and expanding
them in powers of $\Om$ and $k$, we find that
\be
A_x^+(r) \sim O(k), \qquad A_{t0} \sim O(k^2)
 \ee
We thus introduce the following
  \be
 A_x^+(r) = k \le(1 - \frac{a_x(r)}{a_x(\infty)}\ri), \qquad A_{t1} (r) = k^2 a_{t1}(r),
 \ee
 Here by construction $A_x^+$ is normalizable, and from \eqref{w5} we find that $a_x(r)$ satisfies the homogenous equation
  \be \label{p3}
 {1 \ov \sqrt{-g}} \p_r \le(\sqrt{-g} g^{rr} g^{xx} \p_r a_x \ri) -
  \mu^2 (\al-1)^2 g^{tt} g^{xx} a_x  = 0.
 \ee
There is a forced radial equation for the correction to $A_{t1}$ profile $a_{t1}(r)$, but we will not need to solve it explicitly to find the dispersion, analogous to the antiferromagnetic case where the correction to the pion profile was not explicitly needed.
 \be \label{p4}
 {1 \ov \sqrt{-g}} \p_r \le(\sqrt{-g} g^{rr} g^{tt} \p_r a_{t1} \ri) =
  \mu (\al-1) g^{tt} g^{xx} a_x \ .
  \ee
  It is of course critical to note that an infalling and normalizable solution to this equation can always be found.

  We now impose the constraint arising from the conservation of current \eqref{nonAbcons}. Again it is simplest to evaluate it at the AdS boundary first; here the right-hand side of \eqref{nonAbcons} vanishes (as by construction we are looking at a normalizable solution with $A_t^+(\infty) = 0$). We find the dispersion relation
 \be \label{disprel}
 \Om = \om - \mu  = \left(\frac{1}{J^{3t}}\lim_{r\to\infty} \frac{\sqrt{-g}g^{rr}g^{xx}\partial_r a_x}{a_x(\infty)}\right)k^2
 \ee
 This is the desired dispersion relation.

 Let us briefly understand the physical origin of the differences from the linear and gapless dispersion found in the antiferromagnetic case. The dispersion is not linear because the background value of $J^{3t}$ is finite in the $\om \to 0$ limit, unlike in the antiferromagnetic case where it is proportional to $\om$: this means that we are now balancing a term of order $\om$ against one of order $k^2$, rather than a term of order $\om^2$. The mode is not gapless because at infinity the gauge-covariant derivative $d_t = \partial_t + iA_t^3 \to -i\om + i\mu$, resulting in a shift in $\om$. These considerations lead us to expect that if one were able to create a situation with a nonzero $J^{3t}$ in the absence of a background chemical potential -- i.e. a true spontaneous ferromagnet -- one would find precisely the gapless quadratic dispersion of the standard ferromagnetic magnon.

 Note that we can interpret the expression \eqref{disprel} in terms of field theory quantities. Again using expressions from ~\cite{Iqbal:2008by}, we can rewrite the ratio of $\partial_r a_x$ to $a_x$ in terms of a field theory correlator to find
\be
\Om = \left(\frac{G^R(\om = \mu, k = 0)}{J^t_3}\right)k^2,
\ee
with
\be
 G^R (\om, k) =\vev{j^+_x (\om,k) j^-_x(-\om,k)}_{\rm retarded}
\ee
i.e. $G^R_{xx}$ is the field theory retarded correlator for $j_x^+$ evaluated at the nonzero frequency $\om = \mu$.
The prefactor of the quadratic dispersion is consistent with the expected result for a ferromagnet \cite{Read95}.

We now evaluate the constraint \eqref{nonAbcons} at arbitrary $r$. Now the right-hand side no longer vanishes, and we find
\be
\Om = \om - \mu = \ga k^2,
\ee
where
\bwt
\be
\ga = - {1 \ov J^{3t}} \le(\mu (\al-1) \sqrt{-g} g^{rr} g^{tt} \p_r a_{t1}(r) -
\frac{\sqrt{-g} g^{rr} g^{xx} \p_r a_x(r)}{a_x(\infty)} \ri) + a_{t1}(r)  \ .
\ee
\ewt
It can readily checked from~\eqref{p3} and~\eqref{p4} that $\ga$ is independent of $r$, and evaluated at $r \to \infty$ we find that $\ga$ reduces to the expression in \eqref{disprel}. Again, even though we did not need to explicitly solve for $a_{t1}(r)$ to find the dispersion, its fluctuations are essential to make sure that the constraint from non-Abelian current conversation is upheld at all points in the bulk.

\subsection{Evaluation of dispersion}
We now turn to the evaluation of the dispersion $\ga$. This bears a large formal similarity to the evaluation of conductivities studied in \cite{Iqbal:2008by}. We consider the following ``transport coefficient'' defined at all values of $r$:
\be
\sigma(r;\mu) 
 = -\frac{\sqrt{-g}g^{rr}g^{xx}\partial_r a_x (r)}{i\mu a_x(r)}
\ee
The motivation behind this name will soon be made clear. $Q$ is simply related to the boundary value of this object:
\be
\ga = -\frac{i \sigma(r \to \infty;\mu)}{\Xi}
\ee
where we have used~\eqref{susC}. From~\eqref{p3}, we find that $\sig$ obeys a simple radial evolution equation,
\be
\partial_r \sigma = i\mu\sqrt{\frac{g_{rr}}{-g_{tt}}}\left(\frac{\sigma^2}{\Sigma} - \Sigma(\alpha - 1)^2\right),
\ee
with
 \be
\Sigma(r) \equiv \sqrt{\frac{-g}{-g_{rr}g_{tt}}}g^{xx}(r)
\ee
and at the horizon the value of $\sigma$ is fixed by the infalling boundary condition to be
$\sigma(r_h; \mu) = \Sigma(r_h)$ (this is explained in detail in \cite{Iqbal:2008by}).

Before considering general $\mu$, let us consider the limit $\mu \to 0$, in which case $\partial_r \sigma = 0$, $\sigma$ becomes constant in $r$, and is in fact equal to the normal (Abelian) DC conductivity $\sigma_{DC}$ of a $U(1)$ current on this background. Thus we find for the ``magnon'' dispersion~\eqref{disprel}
\be \label{diffu}
\om = -ik^2 \frac{\sigma_{DC}}{\Xi}
\ee
This relation is not surprising; when taking $\mu$ to $0$ we are removing all non-Abelian effects and returning to the zero density system, and~\eqref{diffu} simply describes
the diffusion in the longitudinal channel of a $U(1)$ current. The diffusion constant is seen to be $\frac{\sigma_{DC}}{\Xi}$, as is required by the Einstein relation.

\begin{figure}[!h]
\subfigure[$\;$$\Re\ga$]{\includegraphics[width=7cm]{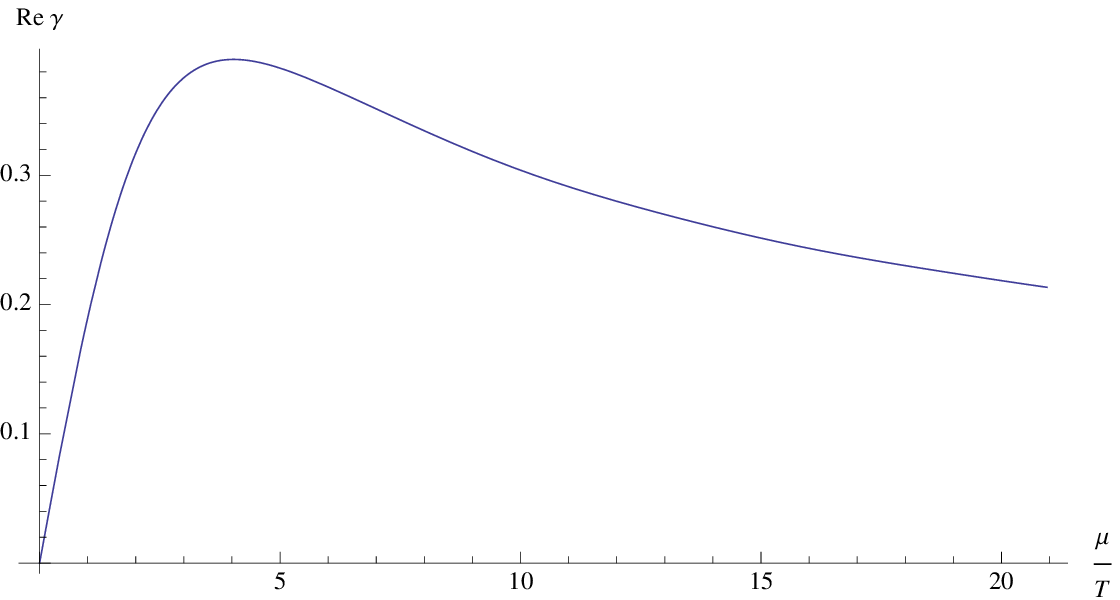}}\hspace{1.5cm}
\subfigure[$\;$$\Im\ga$]{\includegraphics[width=7cm]{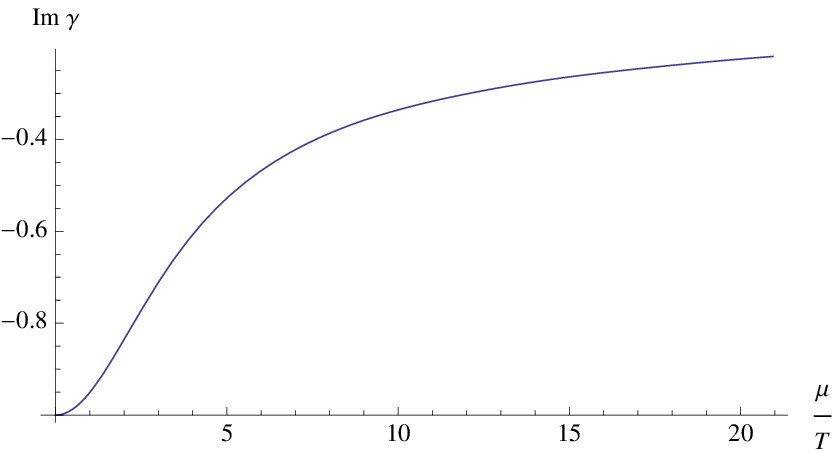}}
\caption{The real and imaginary parts of $\ga$ are shown as a function of $\mu/T$. Note as $\mu \to 0$, $\ga \to -i$ as required by the Einstein relation.}
\label{fig:twogammas}
\end{figure}

\begin{figure}[h]
\begin{center}
\includegraphics[scale=0.5]{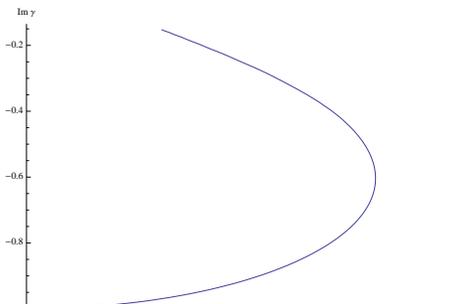}
\end{center}
\vskip -1.0cm
\caption{Movement of $\ga$ in the complex plane as $\mu/T$ is varied from $0$ to $40$.}
\label{fig:Qparamplot}
\end{figure}

 For generic $\mu$ we must evaluate this expression numerically. In this section alone for simplicity we work with the normal AdS-Schwarzschild metric in coordinates with $\eta$ set to $0$. We do not require the RN metric here; even in the absence of a net $U(1)$ charge, one can imagine polarizing thermally excited ``spins'' with an external magnetic field. We do not expect inclusion of a $U(1)$ charge to qualitatively change the results. We work in units where both $\sig_{DC}$ and $\Xi$ have been set to $1$. The results of the numerical evaluation are shown in Figures \ref{fig:twogammas} and \ref{fig:Qparamplot} as a function of $\mu/T$.

Perhaps the most obvious feature of these diagrams is the $\ga$ has a large imaginary part, corresponding to a strong dissipation. This is due to the fact that we are evaluating our bulk equations as a perturbation about a nonzero frequency $\om = \mu$; these waves are infalling at the black horizon and result in a large dissipation. We expect that if we were able to construct a true spontaneous ferromagnet with $\mu = 0$ then our bulk equations would be evaluated at $\om = 0$ and $\ga$ would then be both real and gapless.

As mentioned earlier, this model is the normal phase of the holographic $p$-wave superconductor of \cite{Gubser:2008wv}. According to their discussion, at sufficiently large $\frac{\mu}{T}$ (roughly $\frac{\mu}{T} \sim 15.9$ in our units) this system becomes unstable towards a condensate of the vector fields $A^{1,2}_{x,y}$. It would be interesting to understand further such a condensate -- in our language, a persistent spin current -- from the magnetic point of view taken in this paper.

\section{Discussion and conclusions} \label{sec:con}

In this paper we studied holographic phase transitions associated with condensation of a neutral order parameter, both at finite temperature and associated quantum critical point.
We also considered the embedding of the neutral order parameter into the staggered magnetization for an antiferromagnetic phase. We show that at the macroscopic level one recovers the expected features of the antiferromagnetic phases including existence of two gapless spin waves and their dispersion relations.\footnote{As discussed earlier these
features only depend on symmetry breaking pattern and not details of the microscopic theory.} A similar discussion in a forced ferromagnetic phase reveals
spin waves with quadratic dispersion relations which again agrees with field theoretical expectations.

Our discussions left some loose ends which should be studied further.
The first is the phase diagram for the system in the alternative quantization, for which we only gave a qualitative picture. The phase boundary for
the low temperature stable condensed phase should be mapped out more precisely.
In particular it should be better understood what happens to the system
at $m^2 R^2 = -{27 \ov 16}$. It seems possible that the various scaling exponents
may not obey mean field behavior there. As mentioned earlier while
the appearance of a special value $m^2 R^2 = -{27 \ov 16}$ has to do with nonlinear
structure of the potential, we expect this value and the qualitative behavior we find may not depend on the details of the potential as far as the next order nonlinear term is given by $\phi^4$ (i.e. dictated by a $Z_2$ symmetry). In particular,
we expect similar phenomena should also appear in the context of holographic superconductors. Another important question is to work out the solution for
backreacted spacetime geometry. While we have
shown that deep in the IR
one should find an AdS$_2 \times \RR^2$ with a different cosmological constant, working
out the backreacted metric for all spacetime may reveal new qualitative features.
Also it would be interesting to extend our analysis for both antiferromagnetic and
ferromagnetic spin waves to include backreactions or at nonlinear level~\cite{Bhattacharyya:2008jc}.


The investigation of this paper suggests various other interesting directions to explore. We name a few here:

\subsubsection{The nature of the quantum critical point}

It would be very interesting to better understand the nature of the quantum critical point and
quantum phase transitions discussed in this paper. On the gravity side, along with
the example pointed out in~\cite{Jensen:2010ga}, they all have to do with violation of the
BF bound in AdS$_2$ region, whose critical behavior thus falls into the universality class discussed in~\cite{Kaplan:2009kr}, where a critical line (in the sense of a $(0+1)$-dimensional CFT) ends at critical point. To the left of the critical point an IR
scale is generated, exhibiting the BKT scaling behavior. Below this IR scale,
the system flows to different IR fixed points depending on the charge $q$ of the condensate.
For a neutral order parameter the system flows to a fixed point with $z=\infty$, while
to a $(2+1)$-dimensional CFT when $q$ is sufficiently large. That such distinct physical phenomena share similar IR behavior is striking, yet viewed from $(2+1)$-dimensional perspective, they involve completely different phenomena involving complete distinct
fixed points. We have not yet studied in detail the critical behavior from uncondensed side and spatial as well as temporal correlations near these critical points, which should be accessible from our gravity treatment.

Despite our gravity description, a field theoretical understanding of these critical points remains completely obscure. Can these transitions be understood as lying in the Landau paradigm?
What do they suggest regarding the underlying mechanism for the condensation of charged
and neutral order parameters in theories with gravity duals? Can one find explicit field theoretical models
with similar features? We hope to return to these questions in the future.

\subsubsection{Competition between different orders}

From a more phenomenological point of view, more fields can be added to model more physics: e.g. one natural addition is the inclusion of a superconducting order parameter corresponding to a bulk field $\psi$ charged under $U(1)_{\rm charge}$. One could also turn on a potential $V(\phi^a, \psi)$ to model the
interaction between an AFM and superconducting order parameter. For example, the simplest possibility one could consider is
 \be \label{coup}
 V (\phi^a, \psi) = b \, \vec \phi^2 |\psi|^2
 \ee
Note that a negative value of $b$ will lead to mutual enhancement of
both types of instabilities, while a positive value of $b$ will make them compete.
We expect a rich phase structure to arise from such a construction, which may provide
a strongly coupled example for many condensed matter problems in which such a competition
exists\footnote{Work in progress with M.~Mulligan.}.

A different observation is that in the new IR fixed point
for the condensed phase for the neutral scalar field, since $\tilde R_2 <  R_2$, the
IR dimension~\eqref{nuk} for a (different) charged scalar field appears to be smaller than that in the uncondensed phase. This appears to imply that the condensed phase or associated quantum
critical point may enhance the superconducting instability, say making a previously
stable mode unstable or enhancing the transition temperature to the superconducting
order. A
precise understanding of this will require the explicit construction of the backreacted geometry.

\subsubsection{Coupling to fermions}

Finally, our work raises a number of concrete questions regarding the 
properties of fermions on such backgrounds. It will be important to consider coupling a Fermi surface to our system~\footnote{Work in progress with D.~Vegh.}. 
In an antiferromagnetically ordered state, the nature of 
the Fermi surface provides yet another means to characterize
the phase. In particular, the Fermi surface will reconstruct as a result
of the antiferromagnetic order, in a way that is dependent on the size
of the Fermi surface in the paramagnetic phase. For the system we have analyzed here, in which a continuum limit has already been taken, it will be instructive
to see how to take into account such effects. Relatedly, the evolution of the Fermi surface
across an antiferromagnetic quantum critical point has emerged 
as an important characterization of the nature of the quantum
critical point, and it will be interesting to address this issue 
in our system as well.

Another issue related to fermions involves their low-energy spectral properties.
Similarly to the scalar case described above, one expects that the 
IR scaling dimension $\delta$ of the fermion operator will be
{\it different} before and after the condensation of the scalar $\chi$.
In the holographic formulation, $\delta$ controls the spectral behavior
of the fermions; $\delta > 1$ is somewhat similar to a standard 
Fermi liquid, whereas $\delta < 1$ corresponds to non-Fermi liquid 
behavior~\cite{Faulkner09}. Thus we find the tantalizing possibility 
that the condensation of the scalar and thus the existence of
antiferromagnetic order could change the value of $\delta$, perhaps
driving the system between Fermi liquid and non-Fermi liquid regimes.

\vspace{0.2in}   \centerline{\bf{Acknowledgements}} \vspace{0.2in} We thank D.~Anninos, T.~Faulkner, S.~A.~Hartnoll, D.~K.~Hong, G.~Horowitz, M.~Mulligan, K.~Schalm, T.~Senthil, D.~Vegh and J.~Zaanen for valuable discussions. Work supported in part by funds provided by the U.S. Department of Energy
(D.O.E.) under cooperative research agreement DE-FG0205ER41360 and the OJI program, and by
the NSF Grant No. DMR-0706625 and the Robert A. Welch Foundation Grant No. C-1411.

\begin{appendix}

\section{Effect of magnetic field on IR conformal dimension} \label{App:A}
In this appendix we outline the derivation of \eqref{begnF} demonstrating the effect of a background magnetic field on the IR conformal dimension of a charged scalar field.
\subsection{Dyonic black hole} \label{App:Aa}

We will need to consider the effect of the magnetic field on the geometry. Turning on such an external boundary magnetic field for the $U(1)$ current dual to the
bulk gauge $B_M$ in~\eqref{grac}, the corresponding bulk geometry becomes that of a dyonic
black hole with both electric and magnetic charges,
\be \label{dyonbh}
 ds^2 \equiv g_{MN} dx^M dx^N =  {r^2 \ov R^2} (-f dt^2 + d\vec x^2)  + {R^2 \ov r^2} {dr^2 \ov f}
 \ee
 with
 \be \label{bhga8}
 f = 1 + {Q^2+P^2 \ov r^{4}} - {M\ov r^3}
 \ee
 \be \label{chemM}
 \qquad B_t = \mu_B \le(1- {r_0 \ov  r}\ri), \quad B_x = -{P \ov R^4} y, \quad \mu_B \equiv  {Q \ov R^2 r_0} .
 \ee
$r_0$ is the horizon radius determined by the largest positive root of the redshift factor
\be \label{hord}
f(r_0) =0, \qquad \to \qquad M = r_0^3 + {Q^2+P^2\ov r_0} \ .
\ee
The geometry~\eqref{dyonbh} describes the boundary theory at a finite density with the charge density $\rho$, energy density $\ep$,  entropy density $s$, respectively given by
 \bea \label{thqu}
 \rho = 2 { Q  \ov \kappa^2 R^{2} g_F},  \quad \ep ={M \ov \kappa^2 \ R^{4}}, \quad
  s = {2 \pi \ov \kappa^2} \le({r_0 \ov R}\ri)^{2}  \ .
  \eea
 The external magnetic field $H$ and temperature $T$ are
  \be
   H =  {P \ov R^4}, \quad T = {3 r_0 \ov 4 \pi R^2} \le(1 - { Q^2+P^2 \ov 3 r_0^{4}} \ri)
   \ .
 \ee
$\mu_B$ in~\eqref{chemM} corresponds to the chemical potential of the boundary system.

It is convenient to work with  dimensionless quantities by introducing
\be
Q = \mu r_0^2, \qquad P = h r_0^2
\ee
and rescaling coordinates as
\be
r \to r r_0, \quad (t, \vec x) \to {R^2 \ov r_0} (t, \vec x)
\ee
after which equation~\eqref{dyonbh} becomes
\be \label{dyonbh1}
 {ds^2 \ov R^2}  =  {r^2} (-f dt^2 + d\vec x^2)  + {1 \ov r^2} {dr^2 \ov f}
 \ee
 with
 \be \label{bhga3}
 f = 1 + { 3 \eta \ov r^{4}} - {1 + 3 \eta \ov r^3}, \quad B_t = \mu \le(1- {1 \ov  r}\ri), \quad
 B_x = -h y\
 \ee
and
\be
3 \eta \equiv \mu^2 + h^2 \ .
\ee
Setting $h = 0 $ above one recovers the metric~\eqref{bhmetric1}
used in the main text.

We will be interested in the system at zero temperature, for which
 \be \label{onere}
 Q^2 + P^2 = 3 r_0^4 \quad {\rm or} \quad \mu^2 + h^2 = 3
 \ee
and the near horizon region becomes AdS$_2 \times \RR^2$ with curvature radius
\be
R_2 = {R \ov \sqrt{6}} \ .
\ee
Note that in the zero temperature limit, due to conformal invariance of the
underlying vacuum theory the physically relevant quantity
is the dimensionless ratio
\be
b \equiv {H \ov \mu_B^2}  = {h \ov \mu^2} \ . \label{bdef}
\ee
Using the second relation in~\eqref{onere} we can then express bulk quantities like $h$ (and thus $\mu$) in terms of $b$,
 \be \label{bbre}
 h = {\sqrt{1 + 12 b^2} -1 \ov 2 b} \ .
 \ee

\subsection{Scalar operator dimension in the IR}
Now consider a scalar field in AdS$_{4}$ of charge $q$ and mass $m$, with an
action
\be \label{scaA}
S = -\int  d^{4} x \sqrt{-g} \, \le[(D_M \phi)^* D^M \phi + m^2 \phi^* \phi \ri],
\ee
where the gauge-covariant derivative satisfies
\be
D_M \phi = (\p_M - i q B_M ) \phi \ .
\ee
Note that the action~\eqref{scaA} depends on $q$ only through
 \be \label{mme}
 \mu_q \equiv \mu q \ , \quad h_q \equiv h q
 \ee
which are the effective chemical potential and effective magnetic field for a field of charge $q$.

This problem is now similar to a Landau-level analysis from elementary quantum mechanics. After separation of variables using
\be
 \phi = e^{- i \om t + i k x} Y(y) X(r)\  \label{sepscalar},
\ee
we find that the equations of motion can be written as
\bea
	- {1 \ov \sqrt{-g}} \p_r (\sqrt{-g} g^{rr} \p_r X) + \le(-g^{ii} u^2 + m^2 +g^{ii}\lambda^2 \ri) X = 0 \cr
	-\p_{y}^2 Y+ \le( v^2-\lambda^2\ri) Y = 0 \nonumber \\
 \label{sho}
\eea
with
 \be \label{udef}
  v (y) \equiv  k + h_q y , \quad
  u (r) \equiv \sqrt{g_{ii} \ov -g_{tt}} \le(\om + \mu_q \le(1-{1\ov r} \ri) \ri) \ .
\ee
One then finds that
\be \label{defco}
  Y_n(y)= e^{-\xi^2 \ov 2}\, H_n(\xi), \qquad \xi \equiv \sqrt{\left|h_q\right|} \ \le( y + {k  \ov h_q} \ri)
  \ee
with $H_n$ the usual Hermite polynomials. $X (r)$ is a radial profile that satisfies the scalar wave equation with zero magnetic field $h=0$, except that the momentum $k$ on each constant-$r$ slice has been discretized into Landau levels:
\be
 k^2 \to  2 \left|h_q\right| \le(n+ \ha \ri), \qquad n =0,1, \cdots
 \ee
A similar discussion can be applied to the AdS$_2$ region, where one finds that each Landau level has an effective mass given by
 \be
 m_n^2 = m^2 + 2 \left|h_q\right| \le(n+\frac12 \ri) {1 \ov R^2} \ .
 \ee
The rest then follows exactly from the analysis in~\cite{Faulkner09}, and the IR dimension for $\phi$ is given by
\be
\delta_n^{(B)} = \ha + \nu^{(B)}_n
\ee
with
\be
\nu^{(B)}_n = \sqrt{m_n^2 R_2^2 - {\mu^2 q^2 \ov 36} + {1 \ov 4} } \ .
\ee

Let us examine a bit more closely the $n=0$ mode, which is the most likely to condense,
\be \label{lowM}
\nu^{(B)}_0 = \sqrt{{m^2 R^2 \ov 6} + (6 |bq| -q^2) {\sqrt{1+12b^2}-1 \ov 72 b^2} + {1 \ov 4} }
\ee
where we used~\eqref{bbre} to express $h$ and $\mu$ in terms of the dimensionless boundary quantity $b$ \eqref{bdef}. Also note that $m^2 R^2 = \De (\De -3)$. This is the result \eqref{begnF} in the main text. The critical magnetic field $b_c$ can be found by setting the quantity inside the square root to $0$, and is
\be
b_c=\left|q\right|{D\le[1 + \frac{1}{\sqrt{3}}\sqrt{q^2-2 m^2R^2}\ri] -2 q^2  \ov  D^2 - 12 q^2} \label{bcritical}
\ee
where
\be
D \equiv \le(3 + 2 m^2R^2\ri)
\ee
is a quantity that goes to $0$ when the scalar mass is precisely at the neutral AdS$_2$ BF bound.

It is interesting that the expression in~\eqref{lowM} containing $b$
saturates at a value $|q| \ov {2 \sqrt{3}}$ as $b \to \infty$. Thus if
 \be
 m^2 R^2 + \sqrt{3} |q| < -{3 \ov 2}
 \ee
no matter how large the magnetic field is, a condensate cannot be prevented. This is
surprising and is discussed further in the main text.

\end{appendix}


\begin{thebibliography}{9}
%

\bibitem{Hertz.76}
J.~A.~Hertz,
Phys.~Rev.~B \textbf{14}, 1165--1184 (1976).


\bibitem{Sachdev_book}
S.~Sachdev, \emph{Quantum Phase Transitions}, Cambridge University Press,
Cambridge (1999).


\bibitem{Natphys.08}
{F}ocus issue: Quantum~phase transitions,
Nature~Phys. \textbf{4}, 167--204 (2008).


\bibitem{Gegenwart.08}
P.~Gegenwart, Q.~Si, and F.~Steglich,
Nat.~Phys. \textbf{4}, 186--197 (2008).


\bibitem{Lohneysen.07}
H.~v.~L\"{o}hneysen, A.~Rosch, M.~Vojta, and P.~W\"{o}lfle,
Rev.~Mod.~Phys. \textbf{79}, 1015--1075 (2007).

\bibitem{Si.01}
Q.~Si, S.~Rabello, K.~Ingersent, and J.~L.~Smith,
Nature \textbf{413}, 804--808 (2001).


\bibitem{Senthil.04}
T.~Senthil, A.~Vishwanath, L.~Balents, S.~Sachdev, and M.~P.~A. Fisher,
Science \textbf{303}, 1490--1494 (2004).


\bibitem{AdS/CFT}
J.~M.~Maldacena,
Adv.\ Theor.\ Math.\ Phys.\  {\bf 2}, 231 (1998);
S.~S.~Gubser, I.~R.~Klebanov and A.~M.~Polyakov,
Phys.\ Lett.\ B {\bf 428}, 105 (1998); 
E.~Witten,
Adv.\ Theor.\ Math.\ Phys.\ {\bf 2}, 505 (1998).




\bibitem{Romans:1991nq}
  L.~J.~Romans,
  Nucl.\ Phys.\  B {\bf 383}, 395 (1992)
  arXiv:hep-th/9203018.

\bibitem{Chamblin:1999tk}
  A.~Chamblin, R.~Emparan, C.~V.~Johnson and R.~C.~Myers,
  Phys.\ Rev.\  D {\bf 60}, 064018 (1999)
  arXiv:hep-th/9902170.

\bibitem{quantcritbh}
S.~Sachdev, M.~Mueller,
0810.2005 [cond-mat.str-el]

\bibitem{holographicsc}
  S.~S.~Gubser,
  Phys.\ Rev.\  D {\bf 78}, 065034 (2008)
  [arXiv:0801.2977 [hep-th]];
 \bibitem{hartnolletal}
  S.~A.~Hartnoll, C.~P.~Herzog and G.~T.~Horowitz,
  Phys.\ Rev.\ Lett.\  {\bf 101}, 031601 (2008)
  [arXiv:0803.3295 [hep-th]].

\bibitem{Hartnoll:2009sz}
  S.~A.~Hartnoll,
  Class.\ Quant.\ Grav.\  {\bf 26}, 224002 (2009)
  [arXiv:0903.3246 [hep-th]].
\bibitem{herzogr}
C.~P.~Herzog,
 J.\ Phys.\ A  {\bf 42}, 343001 (2009)
  [arXiv:0904.1975 [hep-th]].
\bibitem{Horowitz:2010gk}
  G.~T.~Horowitz,
  arXiv:1002.1722 [hep-th].



\bibitem{Lee09}
S.-S. Lee, Phys.~Rev.~D \textbf{79}, 086006 (2009).


\bibitem{Liu09}
H.~Liu, J.~McGreevy, and D.~Vegh, arXiv:0903.2477.


\bibitem{Cubrovic09}
M.~Cubrovic, J.~Zaanen, and K.~Schalm,
Science \textbf{325}, 439--444
  (2009).


\bibitem{Faulkner09}
T.~Faulkner, H.~Liu, J.~McGreevy, and D.~Vegh,
arXiv:0907.2694.


\bibitem{Albash:2009wz}
  T.~Albash and C.~V.~Johnson,
  arXiv:0907.5406 [hep-th].

\bibitem{Basu:2009qz}
  P.~Basu, J.~He, A.~Mukherjee and H.~H.~Shieh,
  arXiv:0908.1436 [hep-th].


 \bibitem{soojong}
S.~J.~Rey, Talk at Strings 2007 conference (2007);
Progress of Theoretical Physics Supplement No. 177 (2009) pp. 128-142;
 arXiv:0911.5295 [hep-th].


\bibitem{chenkaowen}
J.~W.~Chen, Y.~J.~Kao and W.~Y.~Wen,
arXiv:0911.2821 [hep-th].


\bibitem{Faulkner:2009am}
T.~Faulkner, G.~T.~Horowitz, J.~McGreevy, M.~M.~Roberts and D.~Vegh,
arXiv:0911.3402 [hep-th].


\bibitem{fabio}
S.~S.~Gubser, F.~D.~Rocha and P.~Talavera,
arXiv:0911.3632 [hep-th].

\bibitem{Denef09}
F.~Denef, S.~A. Hartnoll, and S.~Sachdev,
arXiv:0908.2657.
F.~Denef, S.~A.~Hartnoll and S.~Sachdev,
arXiv:0908.2657 [hep-th],

\bibitem{hofman}
S.~A.~Hartnoll and D.~M.~Hofman,
arXiv:0912.0008 [cond-mat.str-el].


\bibitem{Faulkner:2010tq}
  T.~Faulkner and J.~Polchinski,
  arXiv:1001.5049 [hep-th].

\bibitem{Denef:2009tp}
  F.~Denef and S.~A.~Hartnoll,
  arXiv:0901.1160 [hep-th].


\bibitem{Hartnoll:2009ns}
  S.~A.~Hartnoll, J.~Polchinski, E.~Silverstein and D.~Tong,
  arXiv:0912.1061 [hep-th].

\bibitem{Varma89}
C.~M. Varma, P.~B. Littlewood, S.~Schmitt-Rink, E.~Abrahams,
and A.~E. Ruckenstein,
Phys.~Rev.~Lett. \textbf{63}, 1996 -- 1999 (1989).


\bibitem{Holstein:1973zz}
  T.~Holstein, R.~E.~Norton and P.~Pincus,
  Phys.\ Rev.\  B {\bf 8}, 2649 (1973);
  P.~A.~Lee and N.~Nagaosa, Phys.\ Rev.\  B {\bf 46}, 5621 (1989);
M. Y. Reizer, Phys. Rev. B {\bf 40}, 11571 (1989);
  J.~Polchinski,
  Nucl.\ Phys.\  B {\bf 422}, 617 (1994)   arXiv:cond-mat/9303037;
  C.~Nayak and F.~Wilczek,
  Nucl.\ Phys.\  B {\bf 417}, 359 (1994)
  arXiv:cond-mat/9312086,
  Nucl.\ Phys.\  B {\bf 430}, 534 (1994)
  arXiv:cond-mat/9408016;
  B.~I.~Halperin, P.~A.~Lee and N.~Read,
  Phys.\ Rev.\  B {\bf 47}, 7312 (1993);
Y.~B. Kim, A.~Furusaki, X.-G. Wen, and P.~A. Lee,
Phys.~Rev.~B  \textbf{50}, 17917--17932 (1994);
B.~L.~Altshuler, L.~B.~Ioffe and A.~J.~Millis,
arXiv:cond-mat/9406024;
  S.~S.~Lee, arXiv:0905.4532 [cond-mat].

\bibitem{SachD}
S.~Sachdev, 
arXiv:1002.3823.



\bibitem{bf}
P. Breitenlohner and D. Z. Freedman, Ann. Phys. {\bf 144}, 249 (1982).


  \bibitem{hartlon}
  S.~A.~Hartnoll, C.~P.~Herzog and G.~T.~Horowitz,
  JHEP {\bf 0812}, 015 (2008)
  [arXiv:0810.1563 [hep-th]].


\bibitem{Kaplan:2009kr}
  D.~B.~Kaplan, J.~W.~Lee, D.~T.~Son and M.~A.~Stephanov,
  Phys.\ Rev.\  D {\bf 80}, 125005 (2009)
  [arXiv:0905.4752 [hep-th]].

\bibitem{Jensen:2010ga}
  K.~Jensen, A.~Karch, D.~T.~Son and E.~G.~Thompson,
  arXiv:1002.3159 [hep-th].

\bibitem{FHR}
T.~Faulkner, G.~Horowitz and M.~Roberts, to appear.


\bibitem{Gubser:2009cg}
  S.~S.~Gubser and A.~Nellore,
  Phys.\ Rev.\  D {\bf 80}, 105007 (2009)
  [arXiv:0908.1972 [hep-th]].

\bibitem{Horowitz:2009ij}
  G.~T.~Horowitz and M.~M.~Roberts,
  JHEP {\bf 0911}, 015 (2009)
  [arXiv:0908.3677 [hep-th]].

\bibitem{Gauntlett:2009dn}
J.~P.~Gauntlett, J.~Sonner and T.~Wiseman,
Phys.\ Rev.\ Lett.\ {\bf 109}, 151601 (2009) [arXiv:0907.3796 [hep-th]].

\bibitem{Gauntlett:2009bh}
J.~P.~Gauntlett, J.~Sonner and T.~Wiseman,
JHEP {\bf 02}, 060 (2010) [arXiv:0912.0512 [hep-th]].

\bibitem{Gubser:2009gp}
S.~S.~Gubser, S.~S.~Pufu and F.~D.~Rocha,
Phys.\ Lett.\ B {\bf 683}, 201-204 (2010)
[arXiv:0908.0011[hep-th]

\bibitem{Gubser:2009qm}
S.~S.~Gubser, C.~P.~Herzog, S.~S.~Pufu and T.~Tesleanu,
Phys.\ Rev.\ Lett.\ {\bf 103} 141601 (2009) [arXiv:0907.3510 [hep-th]]

\bibitem{Gubser:2008wz}
S.~S.~Gubser and F.~D.~Rocha, 
Phys.\ Rev.\ Lett.\ {\bf 102} 061601 (2009) [arXiv:0807.1737 [hep-th]]
 
\bibitem{Hartac} We thank Sean Hartnoll
for suggesting this to us.



\bibitem{klebanov:1999}I.~Klebanov and E.~Witten,
 Nucl.\ Phys.\ B {\bf 556} 89 (1999) [arXiv:hep-th/9905104]


\bibitem{Hertog:2004dr}
T.~Hertog and K.~Maeda,
JHEP {\bf 0407}, 051 (2004)
[arXiv:hep-th/0404261]

\bibitem{Hertog:2004bb}
T.~Hertog and K.~Maeda,
Phys.\ Rev.\  D {\bf 71}, 024001 (2005), [arXiv:hep-th/0409314].

\bibitem{Maeda:2009wv}
  K.~Maeda, M.~Natsuume and T.~Okamura,
  Phys.\ Rev.\  D {\bf 79}, 126004 (2009)
  [arXiv:0904.1914 [hep-th]].

\bibitem{Gubser:2005ih}
S.~S.~Gubser, 
Class.\ Quant.\ Grav.\ {\bf 22}, 5121-5144 (2005),
[arXiv:hep-th/0505189].

\bibitem{Iqbal:2008by}
 N.~Iqbal and H.~Liu,
  Phys.\ Rev.\  D {\bf 79}, 025023 (2009)
  [arXiv:0809.3808 [hep-th]].



\bibitem{Gubser:2008wv}
  S.~S.~Gubser and S.~S.~Pufu,
  JHEP {\bf 0811}, 033 (2008)
  [arXiv:0805.2960 [hep-th]].




\bibitem{nabilseandio}D.~Anninos, S.~A.~Hartnoll, N.~Iqbal, in progress.

\bibitem{Bhattacharyya:2008jc}
  S.~Bhattacharyya, V.~E.~Hubeny, S.~Minwalla and M.~Rangamani,
  JHEP {\bf 0802}, 045 (2008)
  [arXiv:0712.2456 [hep-th]].

\bibitem{Auerbach_book}
A.~Auerbach, \emph{Interacting Electrons and Quantum Magnetism},
Springer-Verlag, New York (1994).


\bibitem{Haldane83}
F.~D.~M.~Haldane,
Phys. Rev. Lett. \textbf{50}, 1153 (1983).


\bibitem{Chakravarty89}
S.~Chakravarty, B.~I.~Halperin, and D.~R.~Nelson,
Phys. Rev. B \textbf{39},
2344 (1989).

\bibitem{Chandra90}
P. Chandra, P. Coleman, and A. I. Larkin,
J. Phys.: Condens. Matter {\bf 2}, 7933 (1990).

\bibitem{Read95}
N.~Read and S.~Sachdev,
Phys. Rev. Lett. {\bf 75}, 3509--3512 (1995).


\end{thebibliography}
\end{document}